\documentclass[aps,prb,preprint]{revtex4}
\usepackage{amssymb,latexsym,amsmath}
\usepackage{graphicx}
\usepackage{color}
\usepackage{subfig}
\usepackage{hyperref}

\begin{document}

\title{A Novel Construction of Complex-valued Gaussian Processes with Arbitrary Spectral Densities and its Application to Excitation Energy Transfer }
\author{Xin Chen}
\affiliation{Department of Chemistry,  \\ Massachusetts Institute of Technology, \\ Cambridge, MA 02139}
\author{Jianshu Cao}
\affiliation{Department of Chemistry,  \\ Massachusetts Institute of Technology, \\ Cambridge, MA 02139}
\author{Robert J. Silbey}
\affiliation{Department of Chemistry,  \\ Massachusetts Institute of Technology, \\ Cambridge, MA 02139}
\date{now}
\begin{abstract}

The recent experimental discoveries about excitation energy transfer (EET) in light harvesting antenna (LHA) attract a lot of interest. As an open non-equilibrium quantum system, the EET demands more rigorous theoretical framework to understand the interaction between system and environment and therein the evolution of reduced density matrix. A phonon is often used to model the fluctuating environment and convolutes the reduced quantum system temporarily. In this paper, we propose a novel way to construct complex-valued Gaussian processes to describe thermal quantum phonon bath exactly by converting the convolution of influence functional into the time correlation of complex Gaussian random field. Based on the construction, we propose a rigorous and efficient computational method, the covariance decomposition (CD) and conditional propagation scheme, to simulate the temporarily entangled reduced system.

The new method allows us to study the non-Markovian effect without perturbation under the influence of different spectral densities of the linear system-phonon coupling coefficients. Its application in the study of EET in the Fenna-Matthews-Olson (FMO) model Hamiltonian under four different spectral densities is discussed. Since the scaling of our algorithm is linear due to its Monte Carlo nature, the future application of the method for large LHA systems is attractive. In addition, this method can be used to study the effect of correlated initial condition on the reduced dynamics in the future.
\end{abstract}
\maketitle
\newpage

\section{Introduction}
The study of irreversible open quantum dissipative processes is important in almost every field of condensed-matter physics and chemistry, such as reaction rate theory, ultrafast phenomena, tunneling at defects in solids, and quantum optics, {\it{etc.}}\cite{kuhn,nitzan,scully}.
The 2D spectroscopic experiments \cite{engel} in the light harvesting antennas (LHA) reveal the existence of the long-last coherence in EET. As a result, the real-time dynamics of EEA in the photosynthetic light harvesting environment attracts a lot of theoretical interest and debates\cite{moix}. The almost perfect EET efficiency in the disordered LHA protein environment could be related to the preserved quantum coherence. As an open quantum system, the unique decoherence and relaxation due to the LHA protein environment should play an important role in EET\cite{weiss}.

Dynamic relaxation and decoherence due to environmental fluctuations contains the critical information of  reduced system dynamics and the interaction between system and bath. For classical systems with linear dissipation, Langevin equations (or Ito stochastic differential equation(SDE))\cite{langvine,sde} and its extension provide a simple theoretical (and numerical) framework to describe the interaction between a system and a complex thermal reservoir in terms of stochastic forces and memory friction. For open quantum systems, the corresponding account of quantum noise is still an open question.
Developing a new and rigorous numerical methodology to simulate open quantum dynamics will provide a novel understanding of EET in biological processes.

System bath Hamiltonian with bilinear coupling is the common model to study dissipative dynamics. Particularly, the exciton-phonon coupling Hamiltonian has been used to study exciton transport.
In the Liouville space, the collective motion of environmental phonon modes is reduced to
influence functional, which convolutes the reduced system dynamics. The convolution entangles reduced quantum dynamics. As a result, calculation complexity grows exponentially over the time. The traditional way to avoid the computational issue is using the different truncation schemes based on cumulant expansion or project operator\cite{zwanzig,QNoise,kubocum} in the weak coupling limit. Therefore, the model based on second-order master equations can not accurately describe the interaction of  system and bath and introduce the inconsistence about reduced system dynamics. Under some special cases, the integral equation based on  influence functional can be reduced to a non-perturbative hierarchical equation of motion (HEOM)\cite{kubo1,shao1} and similar ideas have been explored in different contexts\cite{Shapiro1978563}. But the HEOM method is very much limited to the Drude-Lorentzian spectral density ($\it{i.e.}$ exponential kernel.) Therefore, HEOM can not be used for arbitrary spectral densities, requires ad hoc cutoff of the infinite iterative equations and the scaling of the method is nonlinear and unclear. In addition, approaches based on semiclassical path integral\cite{cao1,alan,Stockburger1}, hybrid Ehrenfest and NIBA method\cite{reich}, iterative tensor product method \cite{makri1,makri2} and quantum Brownian motion process\cite{Strunz1996,Shao201029} have been proposed recently and potentially can be used for the general spectral density cases.

The convolution due to influence functional\cite{mak2} makes the computation of reduced system dynamics extremely complicated. Due to the similarity between generalized characteristic function and influence functional, we can map the convolution kernel of influence functional to the covariance matrix of a Gaussian process,
\begin{equation} \label{conv}
\text{Convolution} \longleftrightarrow \text{Correlation}.
\end{equation}
So, we can linearize the computational effort with the covariance decomposition (CD) method to sample the quantum fluctuation of a phonon environment as a complex-valued Gaussian random field.
In other words, constructing Gaussian random field with the CD method can deconvolute the reduced system dynamics. This method will give us a general computational tool to study the effect of phonon environment on  reduced system dynamics in a rigorous and complete way. The sampling strategy for complex-valued Gaussian random process of arbitrary temperature will be more complicated than the real Gaussian random process at the high temperature limit. We will show how to construct the complex-valued Gaussian random process in this paper. However, I will present the computational results for the high temperature limit, $\it{i.e.}$, ignoring the imaginary part of the kernel and quantum detailed balance\cite{cao}. The sampling strategy for the complex-valued Gaussian random process based on the complex unitary transformation will be the future work.

The coupling part of system bath Hamiltonian is critical in determining the interaction between system and environment.
Since the coupling is  bilinear  in the exciton-phonon coupling Hamiltonian, the spectral density of the coupling strength between the system and bath is the major factor in relaxation and decoherence processes\cite{cheng2}. The CD method will be a computationally efficient tool to study the effect of the arbitrary distribution (spectral density) of the linear coupling coefficients on the evolution of reduced density matrix.
Two major challenges from the interaction in the current research work are : 1. the quantum memory effect of  phonon environment(Non-Markovianity); and 2. the correlated initial conditions (inseparability of the system and bath).
The CD method will allow us to address the two challenges in the same framework. Currently, the CD method is base on the influence functional formalism so that the original assumptions of influence functional will be kept
\cite{cal}. Some authors \cite{cao1, shao2,shao3,shao4} have used different approaches to construct to complex-valued Gaussian processes. But the approach we offer here is based on multivariate complex-valued normal distribution function and can take advantage of the established numerical Monte Carlo methods based on  normal distribution functions. Our approach is indifferent to the kernel of influence functional and the choice of spectral density. For example, you can only generate a Markov chain for a Gaussian process with a exponential kernel (Gauss-Markov model) corresponding to the high temperature limit of Drude spectral density. For the general spectral density, it is impossible to generate a Markov chain to sample the corresponding Gaussian process. So our method is general and efficient.

In the discussion section, we like to find out how the shape, such as slope, tail , and center of spectral densities can change the relaxation and coherence of reduced system dynamics.
It is particularly interesting to examine the optical phonon band that have different spectral density from the acoustic one \cite{opticalphonon}. We will look at the geometric impact of the spectral density in this paper.

The paper is organized into five sections: 1. in Sec.~\ref{influe}, we briefly review coherent state path integral and influence functional formalism; 2. in Sec.~\ref{ranop}, we introduce the generalized characteristic function and  random evolution operator. With them, we drive the stochastic integral and differential equations. In addition, we discuss how to derive the Gauss-Markov model in our framework at the high temperature limit; 3. in Sec.~\ref{methodology}, we introduce the CD method and conditional propagation scheme; 4. in Sec.~\ref{results}, we present the benchmarking of the conditional propagation scheme according to the Gauss-Markov model. We also show and discuss the results for FMO under the influence of different spectral densities.

To go beyond the influence functional (the bilinear coupling of system and bath) for the reduced system dynamics, the path integral is the last resort, which will allow us to study the nonlinear coupling of the system and bath motions.

\section{Influence Functional}\label{influe}

In this section, we give a brief description of influence functional based on  the following exciton-phonon coupling Hamiltonian\cite{kubo1,kubo2},
\begin{equation} \label{hamiltonian}
H=H_S(\boldsymbol{a}^{\dagger},\boldsymbol{a})+H_I(\boldsymbol{a}^{\dagger},\boldsymbol{a},X)+H_B,
\end{equation}
where $H_S(a^{\dagger},a)= \sum_k \omega_k\;a^{\dagger}_k\;a_k$, where $\boldsymbol{a}^\dagger$ and $\boldsymbol{a}$ are the set of $a^{\dagger}_k$ and $a_k$, $H_I= V (\boldsymbol{a}^{\dagger}, \boldsymbol{a} )\times X$, $X=\sum_j c_j x_j$ and $H_B=\sum_j \left( \frac{p_i^2}{2m_j}+\frac{1}{2}m_j\omega_j^2 x_j^2 \right)$.
The interaction between system and bath are bilinear which decides that there is only one quanta of energy can be move in or out of the system every time.

The evolution of the isolated
system density matrix $\rho_S(t_f)$ under the Hamiltonian of system  $H_S(\boldsymbol{a}^\dagger,\boldsymbol{a})$ in the path integral representation can be expressed,
\begin{eqnarray}\label{denmat_iso}
& &\rho_S (z_f^*, z'_f,t_f)  = \vert \Psi (t_f) \rangle \langle \Psi'(t_f) \vert = \\ \nonumber
&& \iint N^{-1}d z_f^* d z_f\iint N^{-1} d {z'_f}^* dz'_f \vert z_f \rangle
K (z_f^*,z'_f,t_f,t_i)\langle z'_f \vert,
\end{eqnarray}
where $z$ and $z'$ are complex c-number for bosons and its conjugate, and
the kernel defined as,
\begin{eqnarray}\label{denmat}
K(z_f^*,z'_f, t_f,t_i )  & = & \int D_f [\boldsymbol{Q}(\tau)]
\int D_f [\boldsymbol{Q}'(\tau)] \\ \nonumber
&&\times \exp [(i/\hbar) S_S ( \boldsymbol{Q}, t_f,t_i ) ]
 \exp [- (i/\hbar) S_S^* ( \boldsymbol{Q}', t_f,t_i ) ],
\end{eqnarray}
where
the prime sign $'$ is the indicator of the coordinates associated with the bra state $\langle \Psi ' (t_f) \vert$,
$\int D_f [\boldsymbol{Q}(\tau)]  = \lim_{N\rightarrow\infty}\sum_{i=1}^{N-1} \int N^{-1} dz^*(\tau_i)dz(\tau_i)$,
$\int D_f [\boldsymbol{Q'}(\tau)]  = \lim_{N\rightarrow\infty}\sum_{i=1}^{N-1} \int N^{-1} dz'^*(\tau_i)dz'(\tau_i)$, $N$ is the normalization factor of coherent states.
$S_S ( \boldsymbol{Q}, t_f,t_i )$ and $ S_S^* ( \boldsymbol{Q}', t_f,t_i )$ are actions defined as
\begin{equation}
S_S ( \boldsymbol{Q}, t_f,t_i )=\int_{t_i}^{t_f} d\tau \left( i\hbar \; z^*(\tau)\dot{z}(\tau) -H_S(z^*(\tau),z(\tau))\right),
\end{equation}
and
\begin{equation}
S_S^* ( \boldsymbol{Q}', t_f,t_i )=\int_{t_i}^{t_f} d\tau \left( -i\hbar \; \dot{z}'^*(\tau){z}'(\tau) +H_S(z'^*(\tau),z'(\tau))\right),
\end{equation}
$\boldsymbol{Q}(\tau)$ is the short notation for  the pair of ($z(\tau),z^*(\tau)$), $\boldsymbol{Q}'(\tau)$ for ($z'(\tau),{z'}^*(\tau)$).
Once the bath is coupled to the system, assuming that the initial total density matrix is separable $\rho_{tot}(0)=
\rho_S (0)*\rho_B^e(0)$, where $\rho_b^e$ is the equilibrium density matrix of the bath, $\frac{\exp(-\beta H_B)}{Tr(\exp(-\beta H_B))}$,
the evolution of the reduced system $\rho_S(t_f) = \textbf{Tr} (\rho(t))$ can be expressed in the similar expression in Eq.~\ref{denmat} with the new $K( {z_f}^*, z'_f, t_f,t_i ) $,
\begin{eqnarray}
K( z_f^*, z'_f, t_f,t_i )  & = & \int D_f [\boldsymbol{Q}(\tau)]
\int D_f [{\boldsymbol{Q}'}(\tau)] \\ \nonumber
&&\times \exp [(i/\hbar) S_s (  \boldsymbol{Q}, t_f,t_i ) ] \times F(\boldsymbol{Q},\boldsymbol{Q}';t_f,t_i) \times
 \exp [- (i/\hbar) S_s^* ( {\boldsymbol{Q}'}, t_f,t_i ) ],
\end{eqnarray}
in which $F(\boldsymbol{Q},\boldsymbol{Q}',t_f,t_i)$, influence functional, is defined as,
\begin{eqnarray} \label{eq:inf2}
F(\boldsymbol{Q},\boldsymbol{Q}';t_f,t_i) &=  & \exp  \{   -  \int^{t_f}_{t_i} d\tau \int^{\tau}_{t_i} d\sigma \; (V (\boldsymbol{Q}(\tau)) - V(\boldsymbol{Q}'(\tau))  \\ \nonumber
&&  [\gamma(\tau-\sigma) V(\boldsymbol{Q}(\sigma))-\gamma^\dagger (\tau-\sigma) V(\boldsymbol{Q}'(\sigma))]   \},
\end{eqnarray}
where $\gamma(t)=L_1-i L_2$,
$L_1 (t)=  \int_{0}^{\infty} d\omega J(\omega) \coth (\beta \hbar \omega/2)\cos(\omega t)$, $ L_2 (t)=  \int_{0}^{\infty} d\omega {J(\omega)} \sin(\omega  t)$, and $J(\omega) = \frac{1}{2} \sum_j \frac{c_j^2}{m_j \omega_j} \delta(\omega-\omega_j)$ is the spectral density, which describes the distribution of the coupling strength coefficients between the system and different Harmonic modes.
For the simplification, we assume $\hbar=1$ from now on.

In the next section, we will show how to map the convolution of influence functional to the correlation of Gaussian random process.

\section{General Characteristic Function and Random Evolution Operator}\label{ranop}

Our construction of Gaussian random process mathematically is based on the general characteristics function of classical discrete Gaussian process (DGP). The details of the construction of DGP and general characteristics function (GCF) are briefly reviewed in App.~\ref{dgp}.
By extending this construction, we can obtain the mapping defined in Eq.~\ref{conv}, $\it{i.e.}$, reproducing the convolution of a reduced system quantum dynamics with a Gaussian random process and associated random evolution operator. The goal of this construction is to replace the convoluted path integral defined in Eq.~\ref{denmat_iso} with the path integral conditional on the environment fluctuation.

\subsection{Random Evolution Operator} \label{prop}

In order to map influence functional, we need to extend the real-valued covariance matrix in
Eq.~\ref{covmat} to a complex-valued covariance matrix with kernel, $\gamma(\tau,\sigma)$ in Eq.~\ref{eq:inf2}.
As a result, the complex-valued GCF will be equivalent to influence functional defined in Eq.~\ref{eq:inf2}.
To accommodate the structure of double path integral in Eq.~\ref{denmat_iso}, we propose the following complex-valued
Gaussian stochastic process $\hat{\boldsymbol{\xi}}(t)=[\hat{\xi}(t),\hat{\xi}'(t)]$.
With the stochastic process, the convolution due to the influence functional will be decomposed to a random evolution operator,
\begin{equation}\label{random_op}
F(\boldsymbol{Q},\boldsymbol{Q}';t_f,t_i) = \left \langle \exp \left [ - i \int_{t_i}^{t_f}  d\tau ( V(\boldsymbol{Q}(\tau))\hat{\xi}(\tau)
- V(\boldsymbol{Q}'(\tau))\hat{\xi}'(\tau) ) \right ] \right \rangle_{\hat{\boldsymbol{\xi}}(t)},
\end{equation}
where $\langle \rangle_{\hat{\boldsymbol{\xi}}(t)}$ is the
expectation average over the trajectories $\hat{\boldsymbol{\xi}}(t)$, which are independent of $\boldsymbol{Q}(t)$ and $\boldsymbol{Q}'(t)$, therefore $V(\boldsymbol{Q}(t)$ and $V(\boldsymbol{Q}'(t)$. The detailed derivation can be found in App.~\ref{ifgcf}.
Since the Gaussian random process $\hat{\boldsymbol{\xi}}(t)$ is independent of the system operators, the reduced system density, $\rho_s$ in terms of path integral in Eq.~\ref{denmat_iso}, can be re-written as,
\begin{eqnarray}\label{eq:int-sde}
\rho_s(t_f)
&=& \int D_f [\boldsymbol{Q}(\tau)] \int D_f [{\boldsymbol{Q}'}(\tau)] \\ \nonumber
& &\exp \left [(i/\hbar) S_s (  \boldsymbol{Q}(\tau), t_f,t_i ) \right ] \times F(\boldsymbol{Q},\boldsymbol{Q}';t_f,t_i) \times
 \exp \left [ - (i/\hbar) S_s^* ( {\boldsymbol{Q}'(\tau)}, t_f,t_i ) \right ] \\ \nonumber
 & = & \int D_f [\boldsymbol{Q}(\tau)]\int D_f [{\boldsymbol{Q}'}(\tau)]
\\ \nonumber
&&  \bigg \langle
\exp \left [ (i/\hbar) (S_S (\boldsymbol{Q}(\tau), t_f,t_i ) + V(\boldsymbol{Q}(\tau)) * \xi(\tau) \right ] \label{eq2} \\
&& \exp \left [- (i/\hbar) ( S_S^* ( {\boldsymbol{Q}'(\tau)},  t_f,t_i ) + V(\boldsymbol{Q}'(\tau))* \xi'(\tau) )\right ]
 \bigg \rangle_{\hat{\boldsymbol{\xi}}(\tau)}  \\ \nonumber
 & = & \bigg \langle\mathcal{T}\exp \left \{ - i \int_{t_i}^{t_f}  d\tau [H_{S}+V(\boldsymbol{a}^{\dagger},\boldsymbol{a})\; \hat{\xi} (\tau)] \right \}\rho_S(0) \label{eq3} \\
 & & \mathcal{T} \exp\left \{ i\int_{t_i}^{t_f}  d\tau [H_{S}+V(\boldsymbol{a}^{\dagger},\boldsymbol{a})\;\hat{\xi}'(\tau) ] \right\} \bigg \rangle_{\hat{\boldsymbol{\xi}}(\tau)}  ,
\end{eqnarray}
where $\mathcal{T}$ is time ordering operator. Eq.~\ref{eq3} gives the linear random evolution operators for the forward and backward propagation.
By sampling the trajectories of $\hat{\boldsymbol{\xi}}(t)$ and applying the above equation of motion, we can calculate the evolution of the reduce system density matrix,  $\rho_S(t) = \langle\rho(t\vert \hat{\boldsymbol{\xi}})\rangle_{\hat{\boldsymbol{\xi}}(t)}$. With Eq.~\ref{random_op}, we convert the influence functional convolution to an expectation average of the reduced system dynamics conditional on the Gaussian random field trajectories. The equivalent differential form of Eq.~\ref{eq3} can be expressed as,
\begin{equation}
\label{eq:de-sde}
\frac{d\rho_S(t\vert\hat{\boldsymbol{\xi}})}{dt} = -i\; \boldsymbol{L}_S\;\rho_S(t\vert\hat{\boldsymbol{\xi}}) -i\; [V(a^{\dagger},a)\;\xi(t)\;\rho_S(t\vert\hat{\boldsymbol{\xi}}) - \; \rho_S(t\vert
\hat{\boldsymbol{\xi}})\;V(a^{\dagger},a)\;\xi'(t)],
\end{equation}
where $\boldsymbol{L}_S=[H_S,\; \cdot]$. At the high temperature limit, the two distinct processes, $\xi(t)$ and $\xi'(t)$ will collapsed to one $\xi(t)$ and
the term $ V(a^{\dagger},a)\;\xi(t)\;\rho_S(t) - \rho_S(t)\;V(a^{\dagger},a)\;\xi'(t)$ will become a commutator bracket, $[V(a^{\dagger},a)\;\xi(t),\; \rho(t)]$. Then we can recover the stochastic Liouville equation which is extensively used in the study of the exciton transport\cite{jackson},
\begin{equation}\label{markov-sde}
\frac{d\rho_S(t\vert \hat{\boldsymbol{\xi}})}{dt} = -i\; \boldsymbol{L}(t)\;\rho_S(t\vert\hat{\boldsymbol{\xi}}),
\end{equation}
where $\boldsymbol{L}(t)=[H_s+V(a^{\dagger},a)\;\xi(t),\cdot]$. The details of the reduction of quantum noise to classical noise are discussed in App~\ref{ccgn}.

The structure of influence functional determines that covariance matrix is non-Hermitian, so that it
is different from the one proposed by Miller and Coworkers\cite{miller_cg} for the signal processing and others. The difference is clearly reflected in Eq.~\ref{comp_cov}. Simply speaking, $\xi(t)$ and $\xi'(t)$ are not conjugate to each other, which is the nature of the open quantum dynamics embed in the influence functional.

\subsection{Multichromophore Frenkel-Exciton System}\label{multi}
For the multichromophore Frenkel-Exciton system, we need to consider the path interference. As a result, extra steps should be taken to replicate the right influence functional kernel since besides the time correlation, the correlation between different sites can exist due to path interference. Here, we take a dimer system as an example to explain the extra steps.
The exciton-phonon coupling in a dimer is defined as,
\begin{equation}\label{dimer_ham}
H=H_S+V\times X + I\times H_B,
\end{equation}
where
\begin{equation}
H_S = \left( \begin{array}{cc}
\epsilon_1 & J \\
J & \epsilon_2
\end{array} \right) ,
\end{equation}
\begin{equation}
V= \left( \begin{array}{cc}
V_1 & 0 \\
0 & V_2
\end{array} \right) ,
\end{equation}
$X$ is the bath operator,
and $I$ is identity matrix.
Therefore, in order to replicate the convolution kernel, we need two independent Gaussian
random processes, ${\hat{\boldsymbol{\xi}}}_1(t)$ and ${\hat{\boldsymbol{\xi}}}_2(t)$ with the same kernel. The details of the construction of the Gaussian processes for the dimer Hamiltonian are presented in App.~\ref{mulitlevel}.
The independence means there is no spatial correlation between $\hat{\boldsymbol{\xi}}_1(t)$ and $\hat{\boldsymbol{\xi}}_2(t)$ or the two Gaussian processes have two independent
covariance matrix as defined in Eq.~\ref{comp_cov}. At the high temperature limit, if the phonon band is only coupled to the site energy, the Hamiltonian of dimer in Eq.~\ref{dimer_ham} can be reduced to
the stochastic Hamiltonian,
\begin{equation}
H(t)=
\left( \begin{array}{cc}
\epsilon_1 + {\xi}_1(t) & J \\
J & \epsilon_2  + {\xi}_2(t)
\end{array} \right) ,
\end{equation}
If $V$  has the off-diagonal matrix elements, $V_{ij}$, then spatial correlation will appear between different sites due to the interference between different paths as discussed in the paper of the enhanced coherence\cite{xin}.

\subsection{High Temperature Limit and Gauss-Markov Model}\label{GMM}
Gauss-Markov (Ornstein-Uhlenbeck) model has been used for the study of exciton transport process with memory. The environmental fluctuation is modeled with the Ornstein-Uhlenbeck (OU) process which is a real Gaussian process with exponential kernel. Although Gauss-Markov model is a phenomenological model,  we can derive the Gauss-Markov from  Eq.~\ref{eq:de-sde} with a proper assumption of the spectral density.

If the spectral density is a Drude-form Lorentzian function, $J(\omega)=\frac{\Delta^2}{2\pi}\frac{\beta\gamma\omega}{\gamma^2+\omega^2}$, then $L_1=\Delta^2\frac{\beta\gamma}{2}\exp(-\gamma (\sigma - \tau))$, and $L_2=\Delta^2\exp(-\gamma (\sigma - \tau))$ as discussed in the literature\cite{kubo2, kubo1}.
It can be clearly shown that at the high temperature limit ($\it{i.e.}$, $\beta\rightarrow 0$), $\frac{\beta\gamma}{2}\rightarrow0$ and $L_1 \rightarrow0$. However, $L_2$ has the exponential form. As a result, the Gauss-Markov model can be considered as the high temperature limit of the complex-valued Gaussian process with the Drude-form Lorentzian spectral density.

The Gauss-Markov model \cite{sumi,blumen} has the stochastic Hamiltonian, $H=H_{s}+\xi(t)$, where $\xi(t)$ is the OU process with the kernel $\gamma(\tau-\sigma)=\Delta^2 \exp (\sigma-\tau)$. As a result, $\langle\xi(\tau)\xi(\sigma)\rangle=\langle\xi'(\tau)\xi'(\sigma)\rangle=\langle\xi'(\tau)\xi(\sigma)\rangle=\langle\xi(\tau)\xi'(\sigma)\rangle={\gamma}(\tau-\sigma)$ in Eq~\ref{comp_cov}. In other words, the Gauss-Markov model is a specific case of Eq.~\ref{markov-sde} when the the Gaussian process kernel is exponential. The details of the reduction of complex-valued Gaussian processes to real-valued ones are presented in App.~\ref{ccgn}.

\section{Methodology}\label{methodology}

In this section, we propose the simulation method based on the derivations of Gaussian processes. Firstly, we will discuss the covariance decomposition (CD) method and Monte Carlo sampling strategy to generate the trajectories of environment fluctuations. Secondly, we present the conditional propagation scheme to simulate the reduced system quantum dynamics conditional on the trajectories generated from the CD method.
\subsection{Covariance Decomposition and Monte Carlo Sampling of Environment Fluctuation}\label{samp}
 The Gaussian random process of our construction is independent of the system degree of freedom. Therefore we can sample them before we calculate the reduced system propagation conditional on the Gaussian random process. With the complex-valued PDF in Eq.~\ref{complex_den}, potentially we can sample the discrete complex-valued Gaussian process on the discrete time lattice. At the high temperature, we show in App.~\ref{ccgn}, the complex-valued PDF becomes a real-valued one. For the real-valued PDF, the Cholesky decomposition (CD) \cite{cholesky} is used to sample the discrete Gaussian trajectories by transforming correlated random variables to uncorrelated ones, which decomposes the covariance matrix in Eq.~\ref{ht_cov} into a product of lower and upper triangle matrix, $\Lambda=LL^T$ where $\Lambda=\boldsymbol{\gamma}^{-1}$, the  inverse of covariance matrix $\boldsymbol{\gamma}$, $L$ is a lower triangle matrix. Given the nature of CD, we can parallel the decomposition.

DGP is a multivariate Gaussian random vector.
For a $N$ dimension multivariate Gaussian random vector\cite{MVG,bm}, $\hat{\xi}$, its covariance matrix $\boldsymbol{\gamma}$ is defined as in terms of time correlation functions (kernel of influence functional),
\begin{equation}
\boldsymbol{\gamma}=\langle {\hat{\xi}}{\hat{\xi}}^T \rangle.
\end{equation}
In order to generate the random vector, $\hat{\xi}$ in Eq.~\ref{ranvec} with Cholesky decomposition, we need to generate $N$ independent normal distributed random variables, $\zeta_{i}$,  with Monte Carlo according to the normal distribution function,
$\frac{1}{2\pi}\exp(-\frac{\zeta_i^{2}}{2})$. The random vector $\hat{\xi}$ can be defined as, $\hat{\xi}=L*\hat{\zeta}$,
where $\hat{\zeta}=[\zeta_0,\zeta_1,\cdots,\zeta_N]^T$. With this construction, we can recover the following equality,
\begin{equation}
\langle \hat{\xi}\hat{\xi}^T \rangle = \langle  L\hat{\zeta}\hat{\zeta}^T L^T\rangle = \langle L I L^T \rangle=\boldsymbol{\gamma},
\end{equation}
where $\hat{\zeta}\hat{\zeta}^T=I$ since $\langle \zeta_i \zeta_j \rangle = \delta_{ij}$.
For the complex-valued Gaussian process, since the covariance matrix of PDF in Eq.~\ref{complex_den} is not complex symmetric or Hermitian, finding the reliable algorithm of decomposition is challenging but possible \cite{Willoughby}. We will discuss how to sample a complex-valued Gaussian process in the future paper. So previous work \cite{shao2,shao3} shows the convergence of the complex-valued Gaussian process is slow based their construction. But we anticipate that this construction with tweaks can improve the efficiency of the Monte Carlo for the complex-valued Gaussian process. In this paper, we will limit our method to the high temperature limit and real-valued Gaussian processed as used in Eq.~\ref{markov-sde}.

\subsection{Conditional Propagation Scheme}
Once we can sample the environment fluctuations $\hat{\boldsymbol{\xi}}(t)$ corresponding to influence functional, we can propagate the reduced density matrix according to Eq.~\ref{markov-sde}.
We propose a conditional propagation scheme to calculate the reduced density evolution by averaging over the Gaussian trajectories.
In the propagation scheme, we need to generate one realization of the discrete Gaussian process $\hat{\xi}$ first using the CD method for the whole discrete time grid, $[0, t_1, t_2, \dots, t_{N-1},t]$ as discussed in the previous subsection. We can solve the stochastic Liouville equation in Eq.~\ref{eq:de-sde} by propagating the conditional density matrix using the fourth order Runge-Kutta scheme. We also can propagate the conditional density matrix with the following iterative scheme based on the time-sliced stochastic evolution operator $\exp\left[-i\;dt\; (H_s+V\;\xi(t))\right]$,
\begin{equation}\label{eq:prog}
\rho_S(t+dt\vert \hat{\xi}(t+dt))=\exp\left[-i\;dt\; (H_s+V\;\xi(t))\right]\rho(t\vert\hat{\xi}(t))\exp\left[i\;dt\; (H_s+V\;\xi'(t))\right].
\end{equation}
The steps of the scheme can be described as:
\begin{enumerate}
\item Generate $\hat{\xi}$ using the Monte Carlo according to PDF in Eq.~\ref{complex_den} using the CD method;
\item Propagate the conditional density matrix $\rho_S(t\vert \hat{\xi(t)})$ to $\rho_S(t+dt\vert \hat{\xi}(t+dt))$ using Eq.~\ref{eq:prog} step by step.
\end{enumerate}

From the scheme, we can see that at every step, the density matrix is conditional on the environmental Gaussian random field, $\xi(t)$ and $\xi'(t)$. For the initial density matrix, $\rho_S(0)$ is conditional on the $\xi(0)$ so that we can write it as $\rho_S(0\vert\hat{\xi}(0))$. We choose the left end point, $t$, of the interval $[t,\; t+dt]$ to define our stepwise evolution operator, $\exp\left(-i\;dt\; (H_S+V\;\xi(t))\right)$ and $\exp\left(i\;dt\; (H_S+V\;\xi'(t))\right)$. It might be interesting to explore the numerical scheme using the middle point or higher order approximation in the stepwise evolution operator to have better efficiency and accuracy at larger time step size, $dt$.

\section{Results and Discussion}\label{results}

We introduce the conditional propagation scheme to compute the evolution of conditional reduced density matrix, $\rho_S(t\vert \hat{\xi})$ dependent on Gaussian trajectories, $\xi(t)$. Since the Gauss-Markov model has been used widely before, it can sever as a good benchmark model to validate the propagation scheme.
Besides it, we will apply the scheme on the FMO system to study the dynamic effect of spectral densities.
\subsection{Benchmark with Gauss-Markov Model}\label{benchmark}

The Gauss-Markov Model has been extensively used to study exciton transport\cite{blumen}. We take a symmetric dimer again as an example, which has the following stochastic Hamiltonian,
\begin{equation}
H=\sum_{k=1}^2 \epsilon_k \vert k \rangle\langle k \vert + J( \vert 1  \rangle \langle 2 \vert+ \vert 2  \rangle \langle 1 \vert )+ \delta\epsilon_{1} \vert 1 \rangle \langle  1 \vert+ \delta\epsilon_{2} \vert 2 \rangle \langle  2 \vert,
\end{equation}
where the energy fluctuations, $\delta \epsilon_{i}$, are the Gauss-Markov process with exponential kernel
\begin{equation}
\langle \delta\epsilon_{i}(t)\delta \epsilon_{j}(t')\rangle=\Delta^2 \exp(-\gamma \vert t-t'\vert )\delta_{ij}.
\end{equation}
For the symmetric dimer ($\epsilon_1=\epsilon_2$), the local master equation (partial ordering prescription) of reduced density matrix is  a set of coupled integro-differential equation \cite{palenberg},
\begin{eqnarray} \label{eqm}
\frac{p_s(y)}{dy}&=& \psi_s (y) \\ \nonumber
\frac{\psi_s(y)}{dy} &=& -(1-2\Delta^2_s\;g_1(y)) \; p_s(y) -2\Delta_s^2\; g_2(y) \;\psi_s(y),
\end{eqnarray}
where $P(t)=\rho_{11}-\rho_{22}(t)$, $\psi(t)=i(\rho_{21}(t)-\rho_{12}(t))$, $p_s(y)=P(y/2J)$, $\psi_s(y)=\psi(y/2J)$, $y=2Jt$, $t$ is the time,  and $\Delta_s=\frac{\Delta}{2J}$.
Using the symmetric dimer, we can benchmark our conditional propagation scheme in two ways: 1. the first benchmarking, comparing the results of the conditional propagation scheme with the results of Eq.~\ref{eqm} at the weak damping limit, $\Delta/J \ll 1$; 2. the second benchmarking, comparing the CD method with the Markov chain Monte Carlo (MCMC) sampling method which only works for the exponential kernel\cite{press}.

\begin{figure}
\begin{tabular}{cc}
\subfloat[]{\includegraphics[width=3.2in]{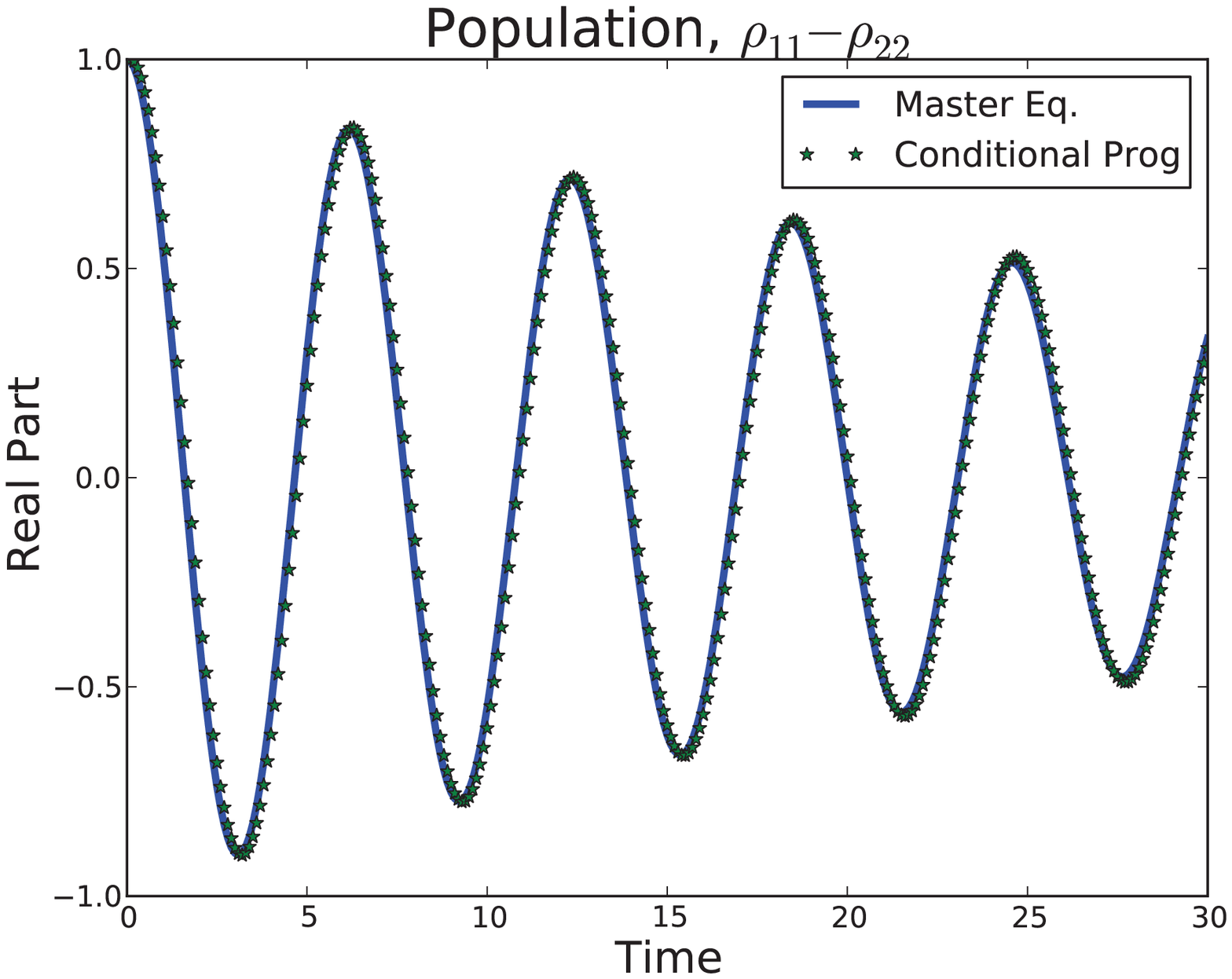}}
   & \subfloat[]{\includegraphics[width=3.2in]{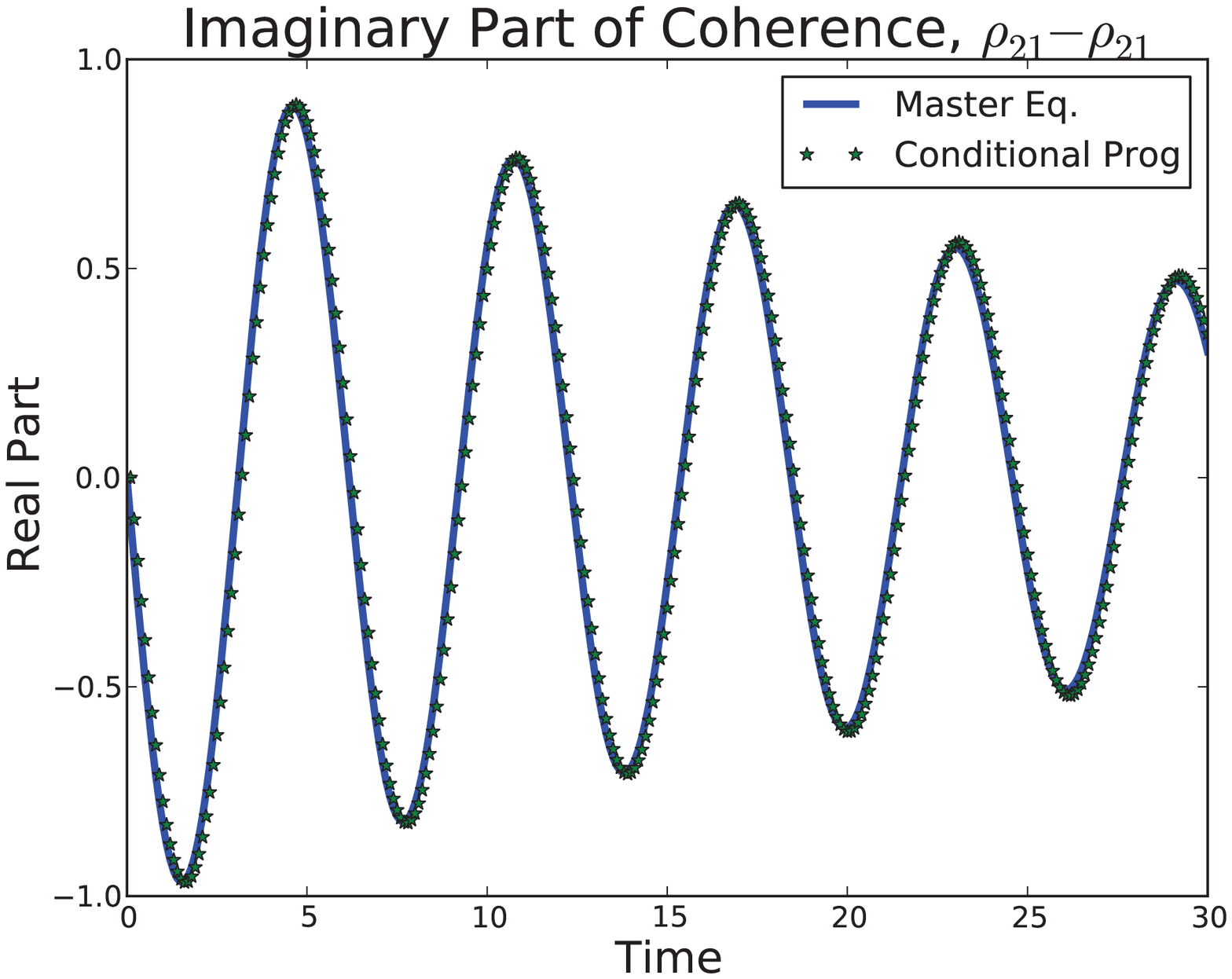}} \\
\end{tabular}
\caption{Comparison of the results of conditional propagation scheme and the POP second-order master equation at the weak damping limit}\label{weakc}
\end{figure}

\begin{figure}
\begin{tabular}{cc}
\subfloat[]{\includegraphics[width=3.2in]{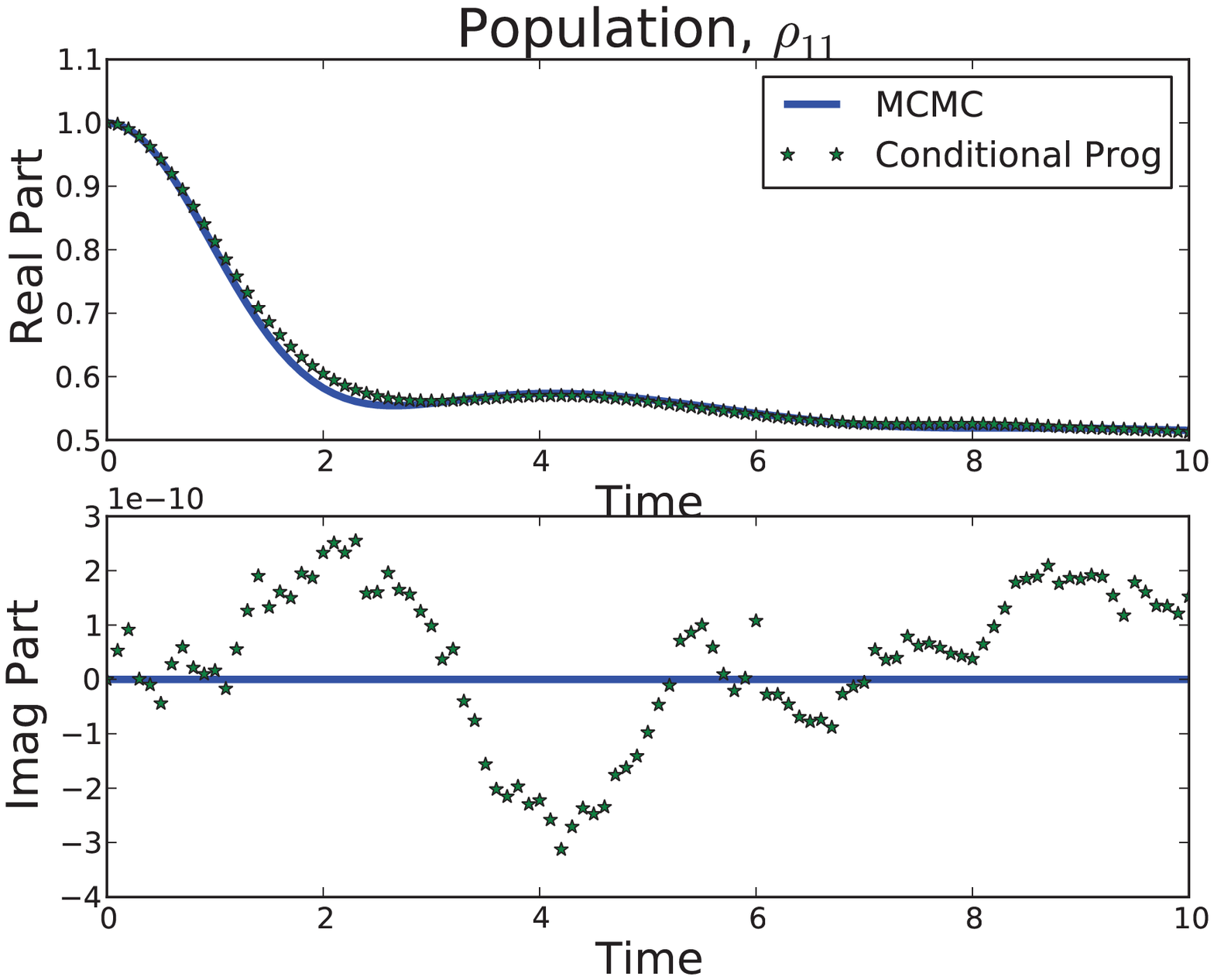}}
 & \subfloat[]{\includegraphics[width=3.2in]{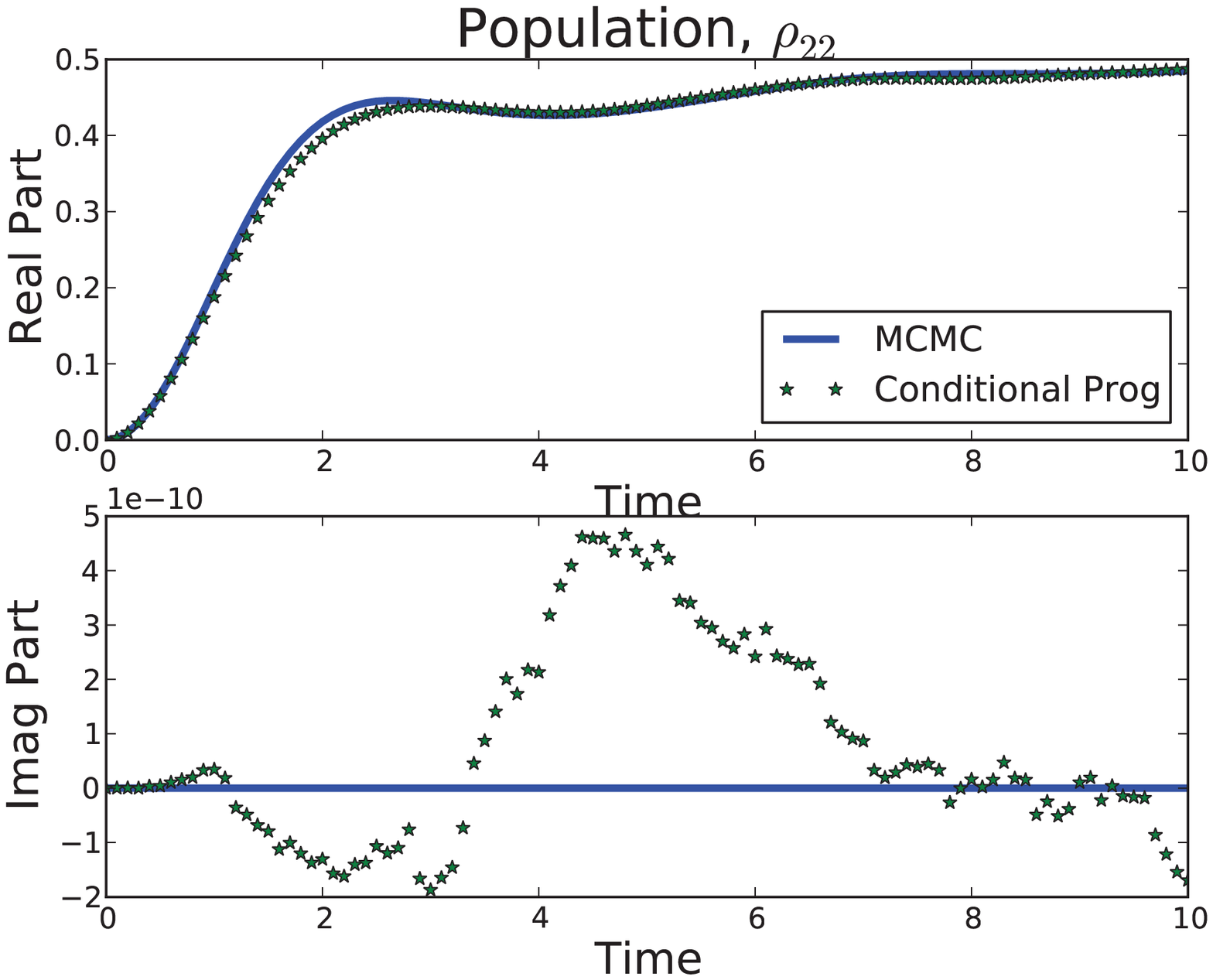}} \\
\end{tabular}
\centering
\subfloat[]{\includegraphics[width=3.2in]{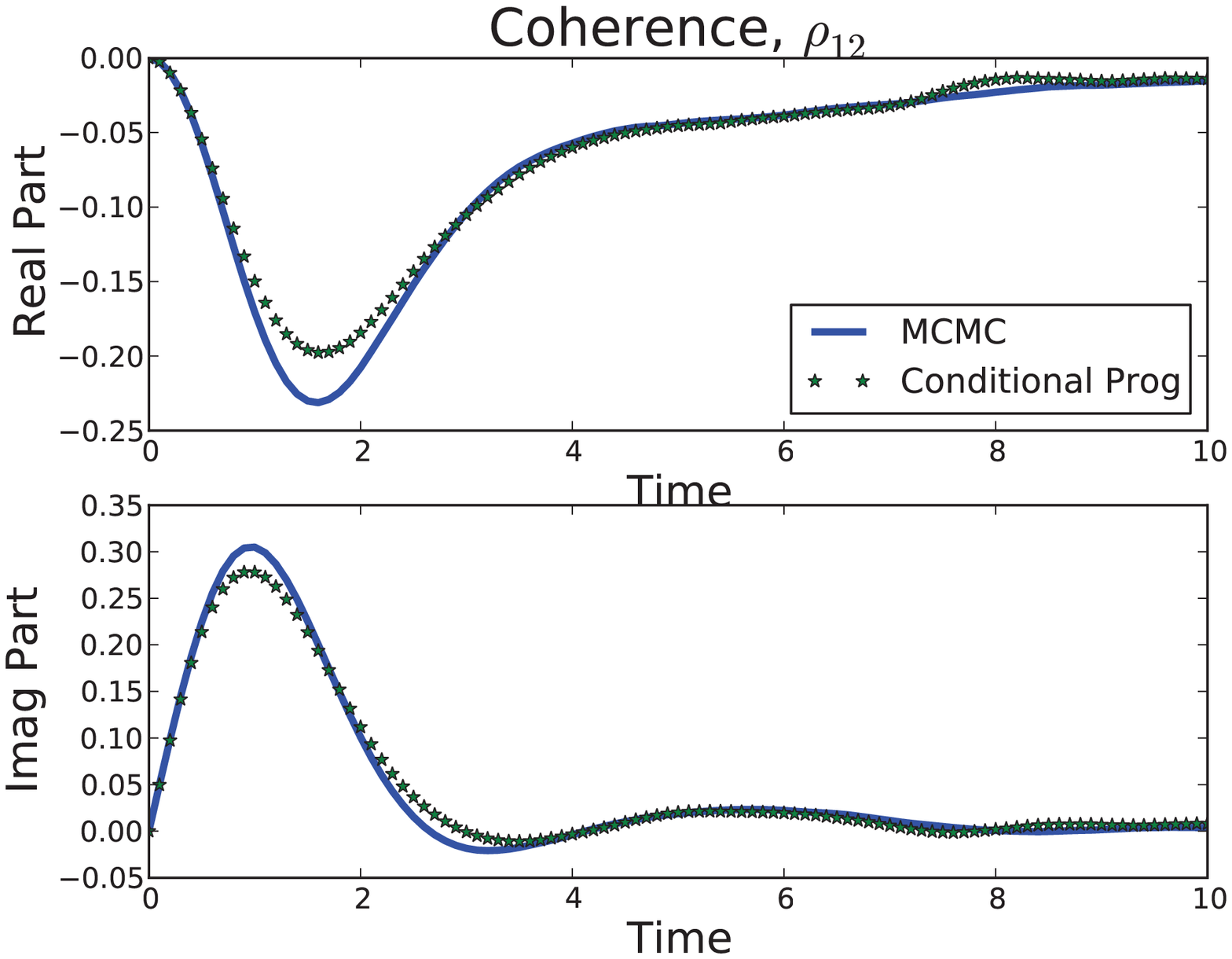}}
\caption{Comparison of the results of the MCMC and CP methods (the imaginary part of the populations $\rho_{11}$ and $\rho_{22}$ in Panels (a) and (b) are at the magnitude of 1.0e-10)}\label{sampc}
\end{figure}

In the calculations, we choose $\epsilon_1=\epsilon_2=0$, $J=0.5$, $\Delta=0.05$, and $\gamma=1.0$.
Figure~\ref{weakc} shows the first benchmarking results for the population $\rho_{11}-\rho_{22}$ and coherence $\rho_{21}-\rho_{12}$. The two results agree with each other very well.
Figure~\ref{sampc} shows the second benchmarking results for the population $\rho_{11}-\rho_{22}$ and coherence $\rho_{21}-\rho_{12}$. In the second benchmarking, the MCMC method is used to generate the independent Gauss-Markov processes $\xi(t)$. We have a good agreement as well. For both benchmarking, we use 5000 trajectories.

\subsection{Effect of Spectral Density in the FMO System}\label{spectral}
Spectral density plays an important role in the dynamics of reduced system since the distribution of coupling coefficient decides how the energy flows into and out of the system. Mathematically, the relaxation process due to stochastic Gaussian environment is governed by the kernel of Gaussian process.
The system of FMO has been studied\cite{Ishizaki,engel} extensively, we take the FMO model as an example to look at the influence of different kernels. The system Hamiltonian in the references\cite{xin2,Hamil1,cho} is used.
The recent work \cite{moix} shows that the coherence in FMO system could depend on the initial condition. However, we in this paper will focus on how dephasing rates
get manipulated by the spectral density given a fixed initial condition at site 1 and thereof coherence. We treat the Hamiltonian more as another model system.

The real part of kernel $\gamma(\tau-\sigma)$ is determined by spectral densities,
\begin{equation}
L_{1}(\tau-\sigma)=\int_{0}^{\infty}d\omega J(\omega)\coth(\beta\hbar\omega/2)\cos(\omega t).
\end{equation}
Ohmic spectral density has the form $J(\omega)=\eta \omega$ and it leads to the $\delta(t)$ spectral density\cite{ingold}.
However, Ohmic spectral Density with  Lorentzian or exponential cutoff  is used extensively in the study of non-Markovian excitation
energy transfer in the light harvesting complex. Ohmic spectral density with  Lorentzian cutoff or Drude spectral density is particularly popular because of its connection to Gauss-Markov process (exponential kernel) as revealed in Subsection~\ref{GMM}.
For the current model, the geometry of spectral density should play an role in the dynamics of reduced systems.
For example, at the high temperature or strong coupling limit, the probability of multiphonon gets high and the phonon side band becomes closer to a Gaussian distribution \cite{holeburning1, holeburning}. The shift will change the reduced dynamics.
The Gaussian spectral density is used to model the optical phonon band\cite{opticalphonon}. It is interesting to  study the effect of a optical Gaussian phonon on EET \cite{Yang-flem}.
It is also important to notice that the different cutoffs of the Ohmic spectral density will generate different non-$\delta$ kernels at hight temperature limit and different non-Markovian effect on the reduced system.

In this section, we choose four different kernels: 1.
the kernel of Drude spectral density, $J(\omega)={\eta\omega} \frac{\gamma}{\pi(\gamma^2+\omega^2)}$ in reference \cite{kubo1}; 2.
the kernel of Ohmic spectral density with  exponential cut-off $J(\omega)={\eta\omega} \frac{1}{2\omega_c} \exp(-\omega/\omega_c) $ with $\omega_c = 5 ps^{-1} $; 3.
the kernel of the first Gaussian spectral density $\eta \omega \frac {1}{\sigma \sqrt{2\pi}} \exp(-\frac{(\omega-\omega_{op})^2}{2\sigma^2})$ with $\omega_{op}=0$ with $\sigma=\gamma$; 4. the kernel of the second Gaussian spectral density with $\sigma=2\gamma$. All the parameters are set up according to the paper \cite{xin}.

\begin{figure}[hb]
\centering
  \includegraphics[width=4in]{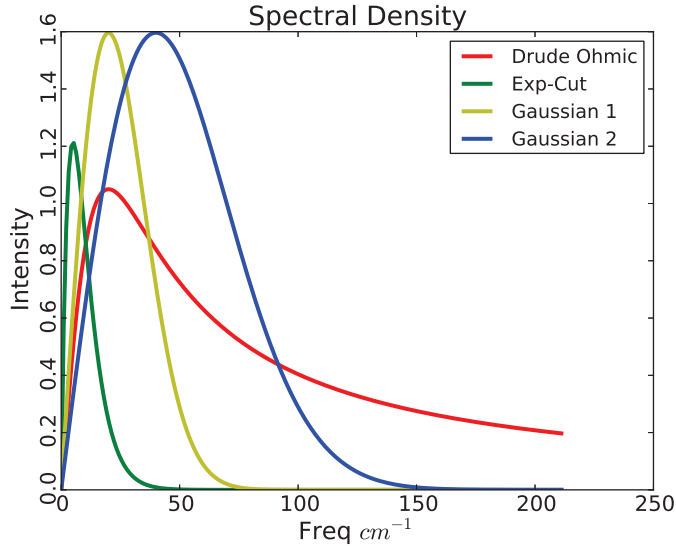}
  \caption{We present four spectral densities: Ohmic spectral density with Lorentzian cutoff, Ohmic spectral density with exponential cutoff, and spectral densities with Gaussian cut off with $\sigma=\gamma$ and $\sigma=2\gamma$ ($\gamma=5\ ps^{-1}$).}\label{spec}
\end{figure}

\begin{figure}
\begin{tabular}{cc}
\subfloat[]{\includegraphics[width=3.2in]{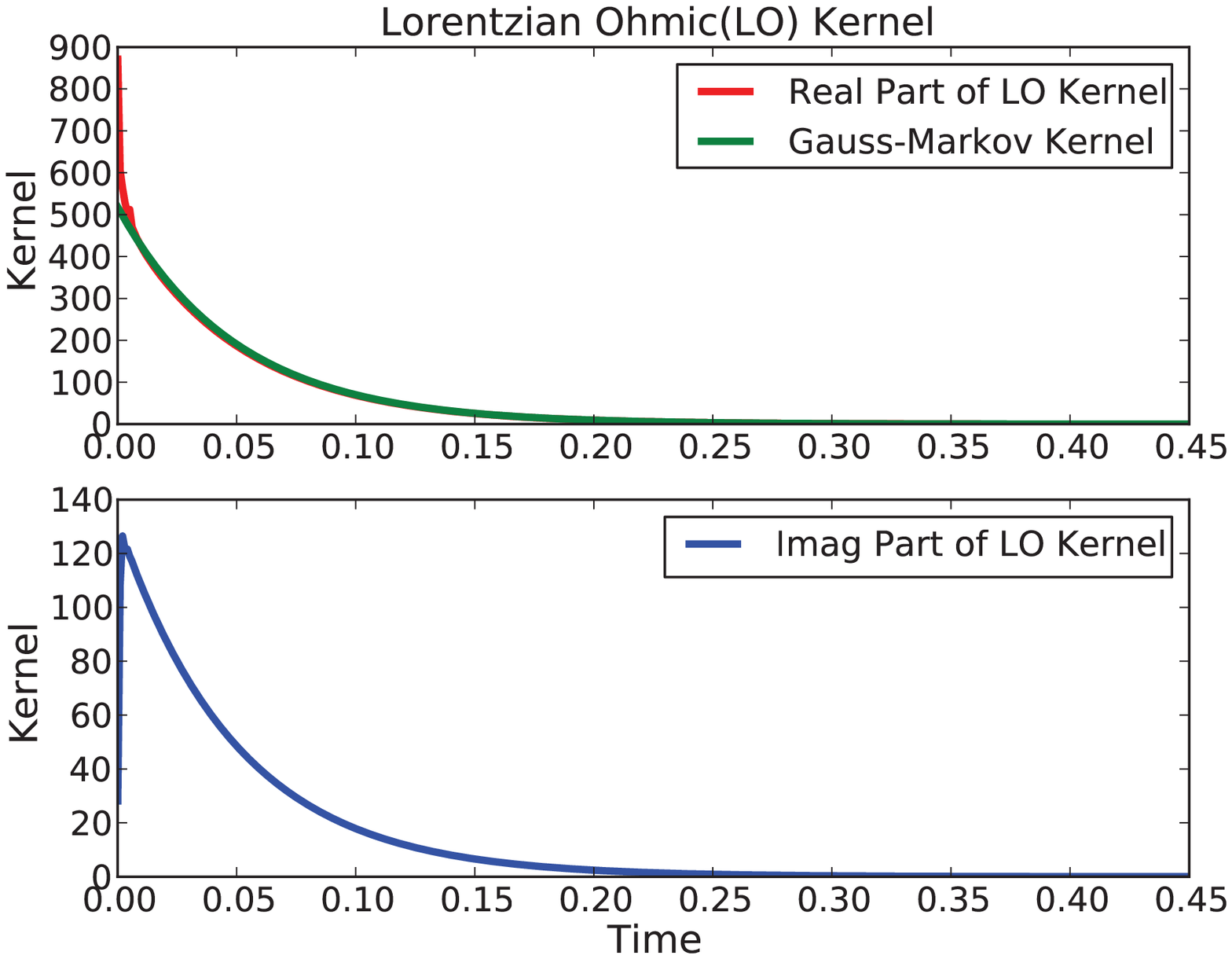}}
   & \subfloat[]{\includegraphics[width=3.2in]{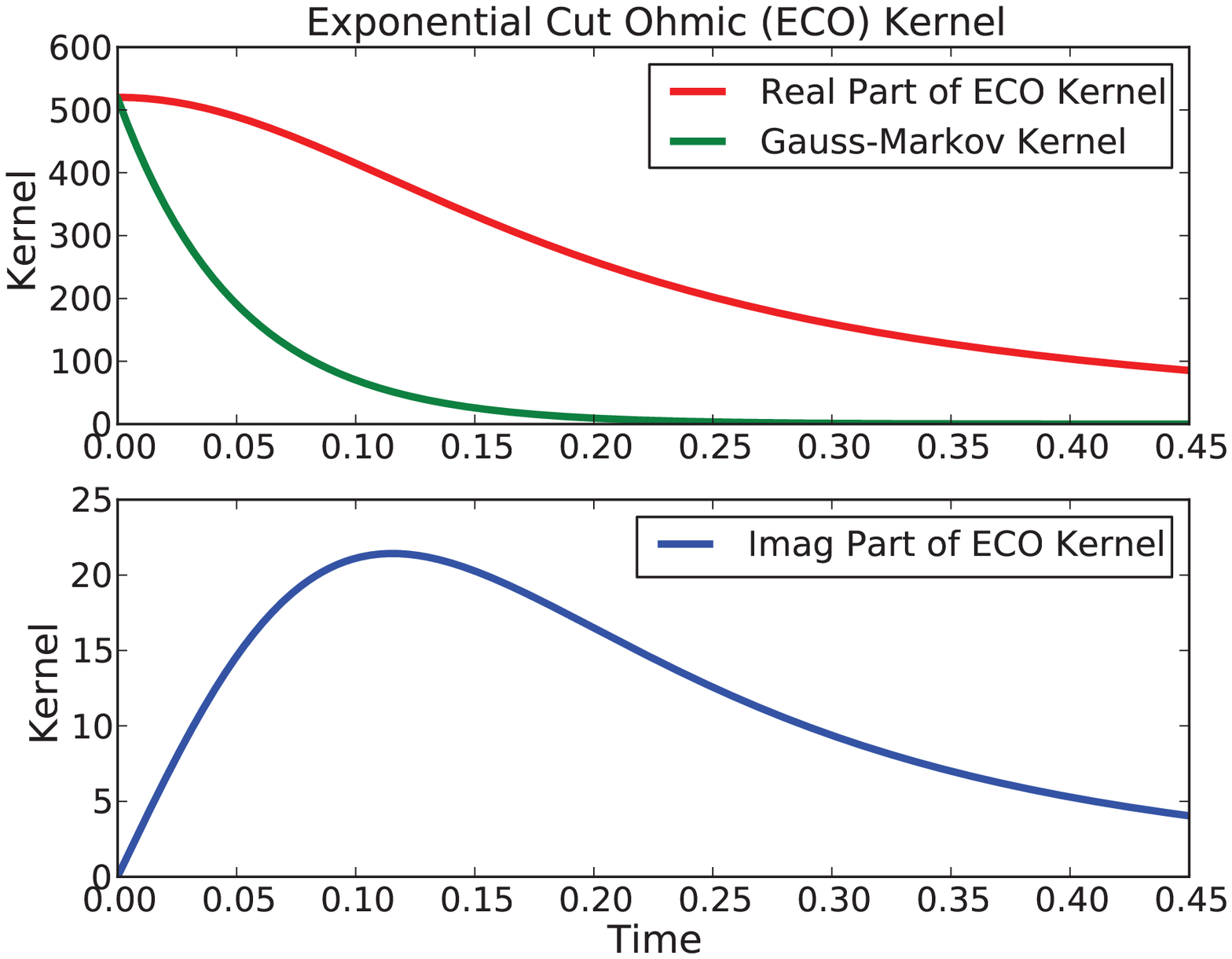}} \\
   \includegraphics[width=3.2in]{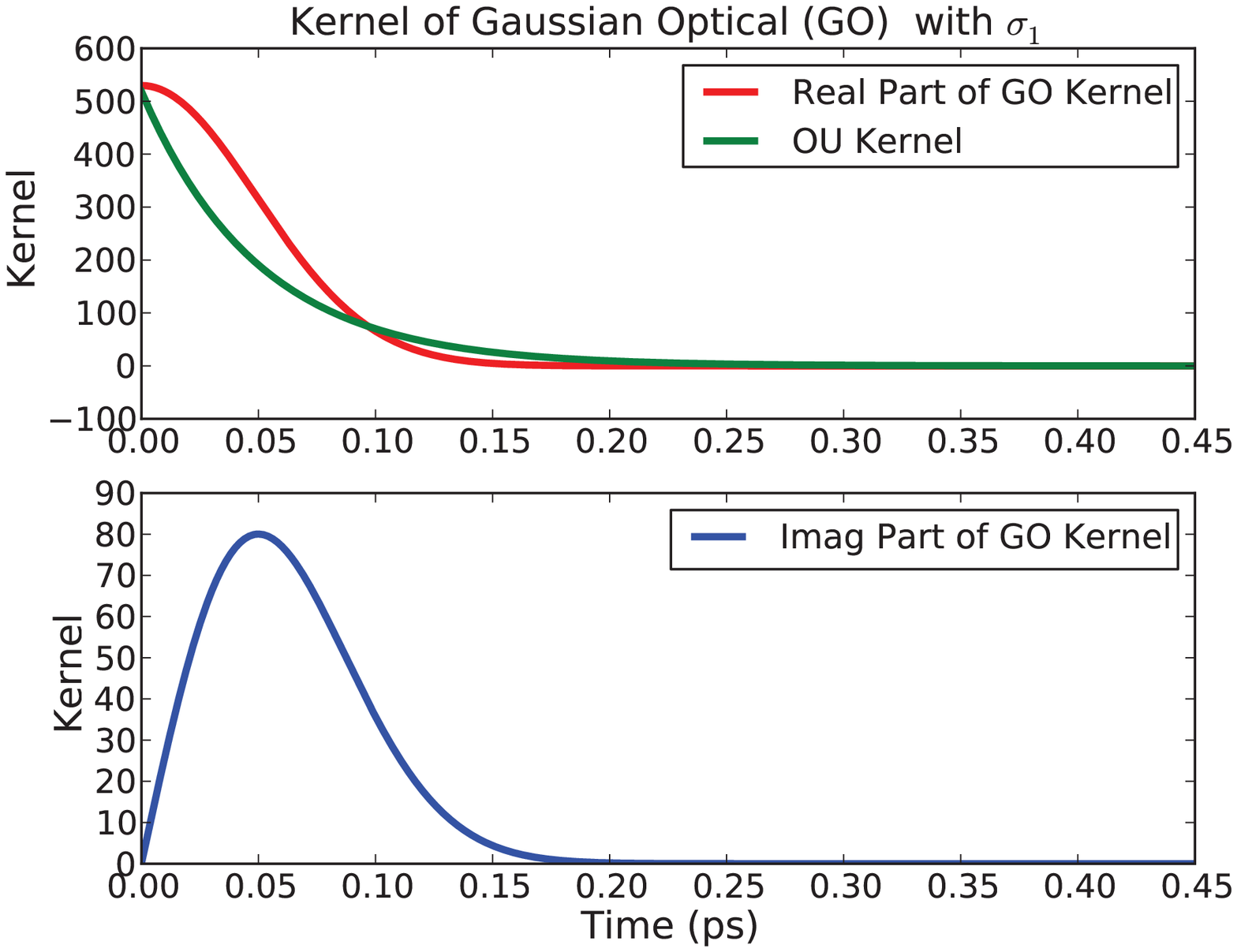}
  & \includegraphics[width=3.2in]{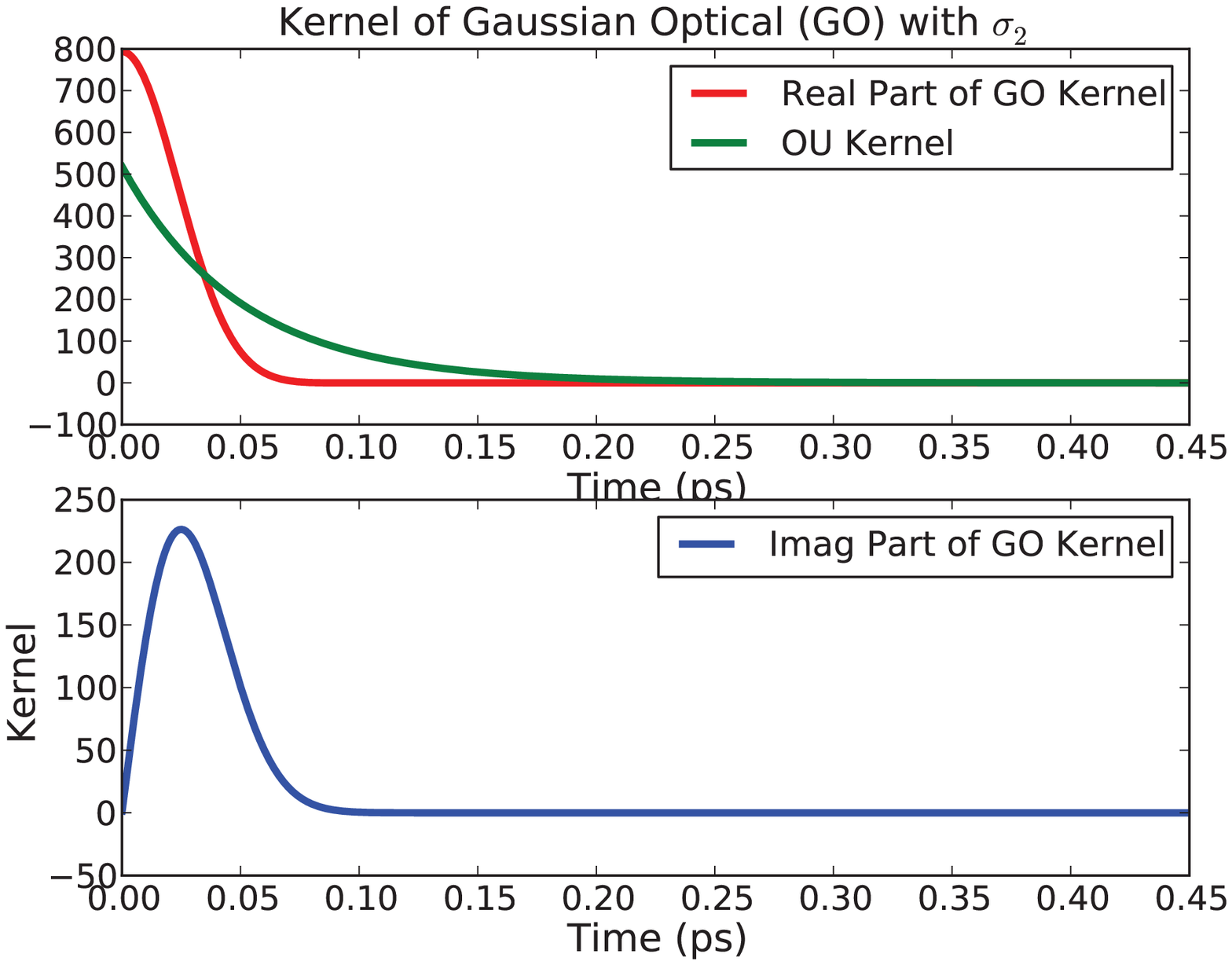} \\
\end{tabular}
\caption{Kernels of four different spectral densities,  Lorentzian (a), Exponential Cut (b), Gaussian with $\sigma=\gamma$ (c) and Gaussian with $\sigma=2\gamma$ (d) ($\gamma=5\ ps^{-1}$)}\label{kernel}
\end{figure}

The Drude spectral density is used as the reference reorganization energy $\eta = \frac{\Delta^2\beta}{2}$. In order to compare results at the same level, we normalize the four spectral densities to the reference reorganization energy, $\int_0^{\infty} d\omega J(\omega)/\omega$ \cite{cheng}.
The superimposing of the four different spectral densities is shown in Figure~\ref{spec}.
In the figure, ohmic with exponential cut has the peak around 15$cm^{-1}$, the Drude and first Gaussian spectral density around 25$cm^{-1}$, the second Gaussian spectral density around 40 $cm^{-1}$. The major difference between the Drude and first Gaussian is the weight of high frequencies. The Drude has longer tail and more weight at the high frequency domain. Ohmic spectral density with exponential cutoff, compared to other spectral density, mostly concentrates in the low frequency domain. The comparisons of the four corresponding kernels are shown in Fig.~\ref{kernel}. Panel (a) in Fig.~\ref{kernel} shows that the real part of the kernel of the Drude spectral density is almost identical to the exponential kernel and tells us that at high temperature, the model with Drude Ohmic spectral density will collapse to the Gauss-Markov Model.
We also can observe that even though the Drude and first Gaussian spectral densities have the same peak, they have different kernels in Fig.~\ref{kernel}. The real part of the kernel of the Drude spectral density has longer curve and bigger area under the curve compared to the first Gaussian spectral density.

\begin{figure}
\begin{tabular}{cc}
\subfloat[]{\includegraphics[width=3in]{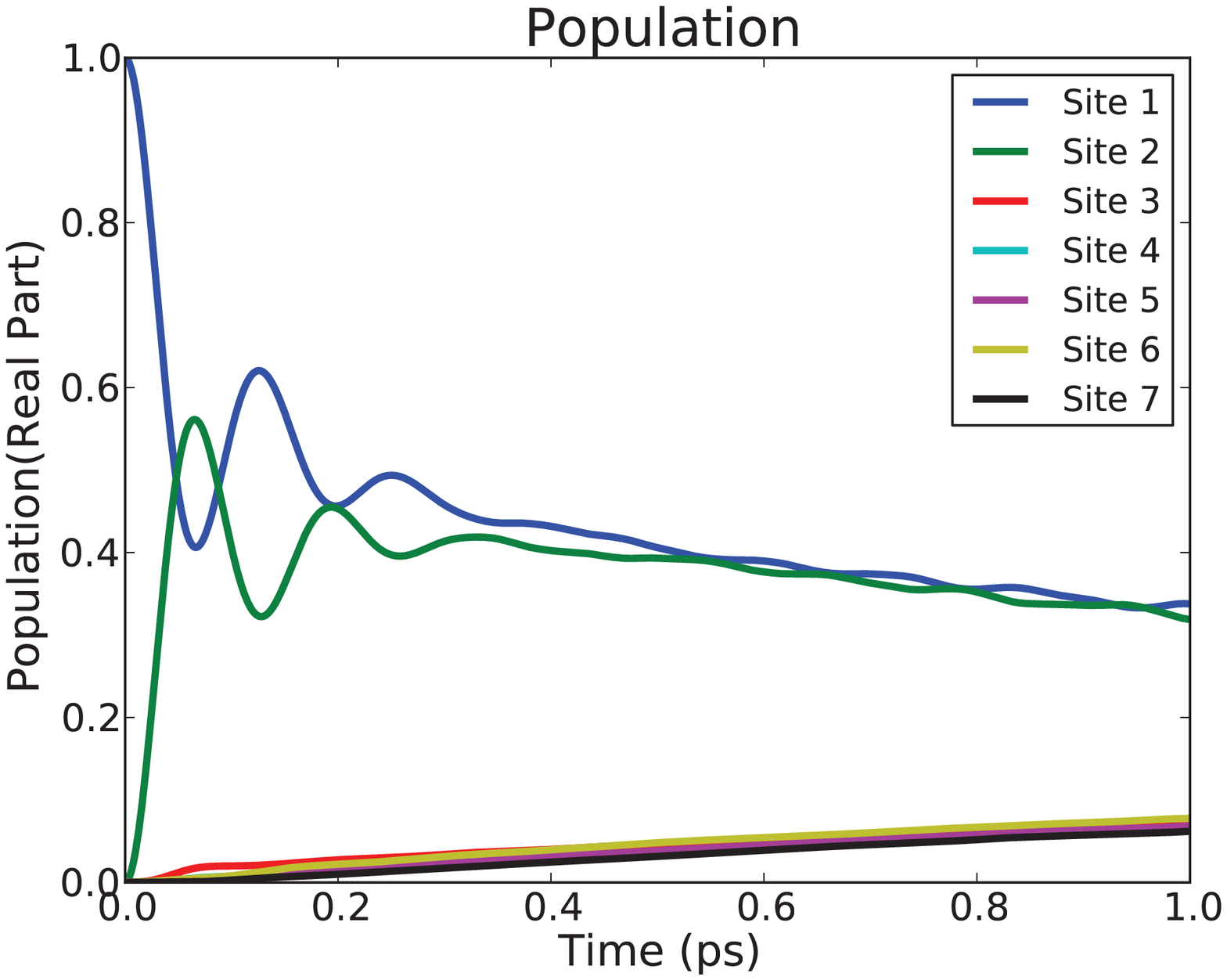}}
   & \subfloat[]{\includegraphics[width=3in]{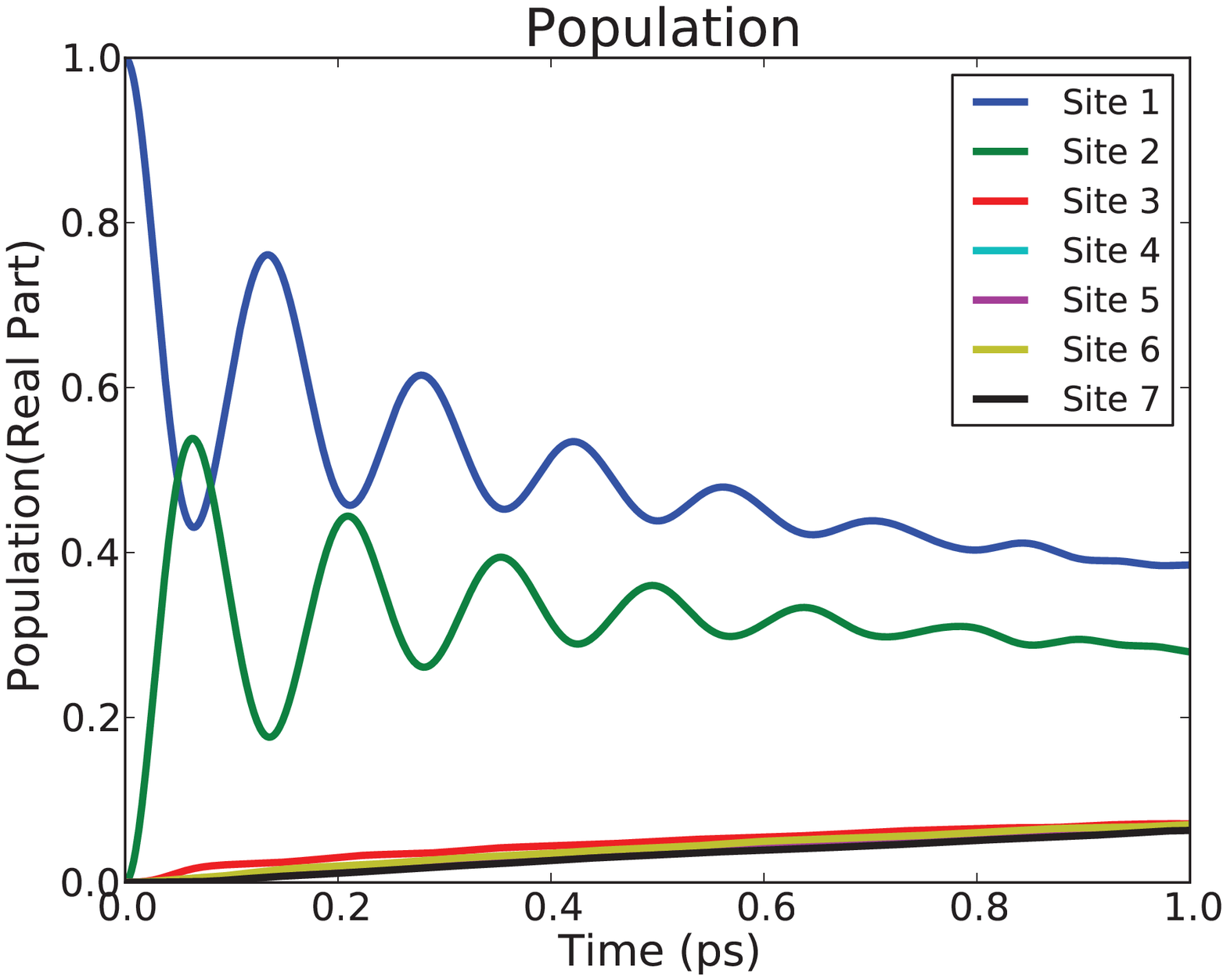}}\\
\subfloat[]{\includegraphics[width=3in]{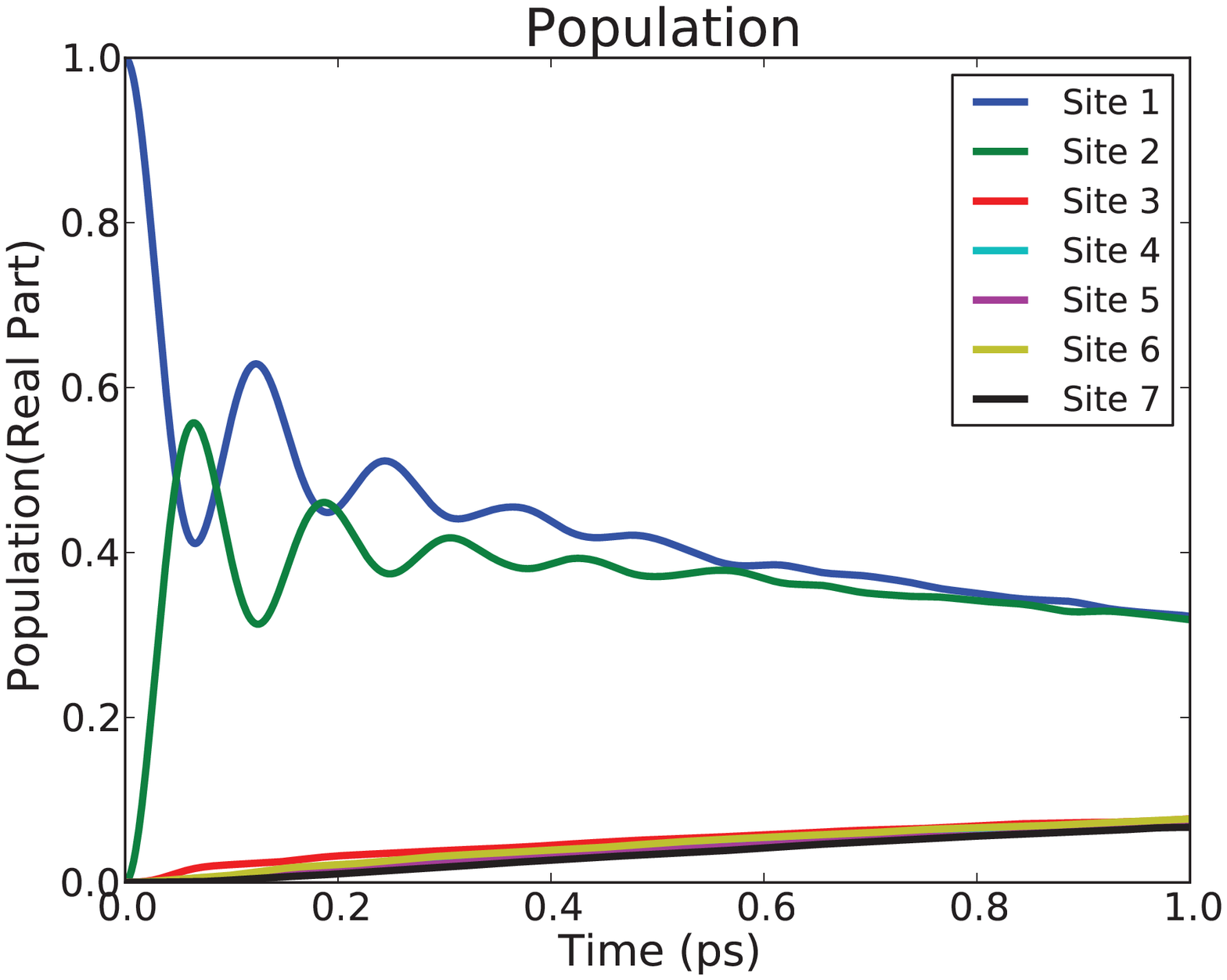}}
   & \subfloat[]{\includegraphics[width=3in]{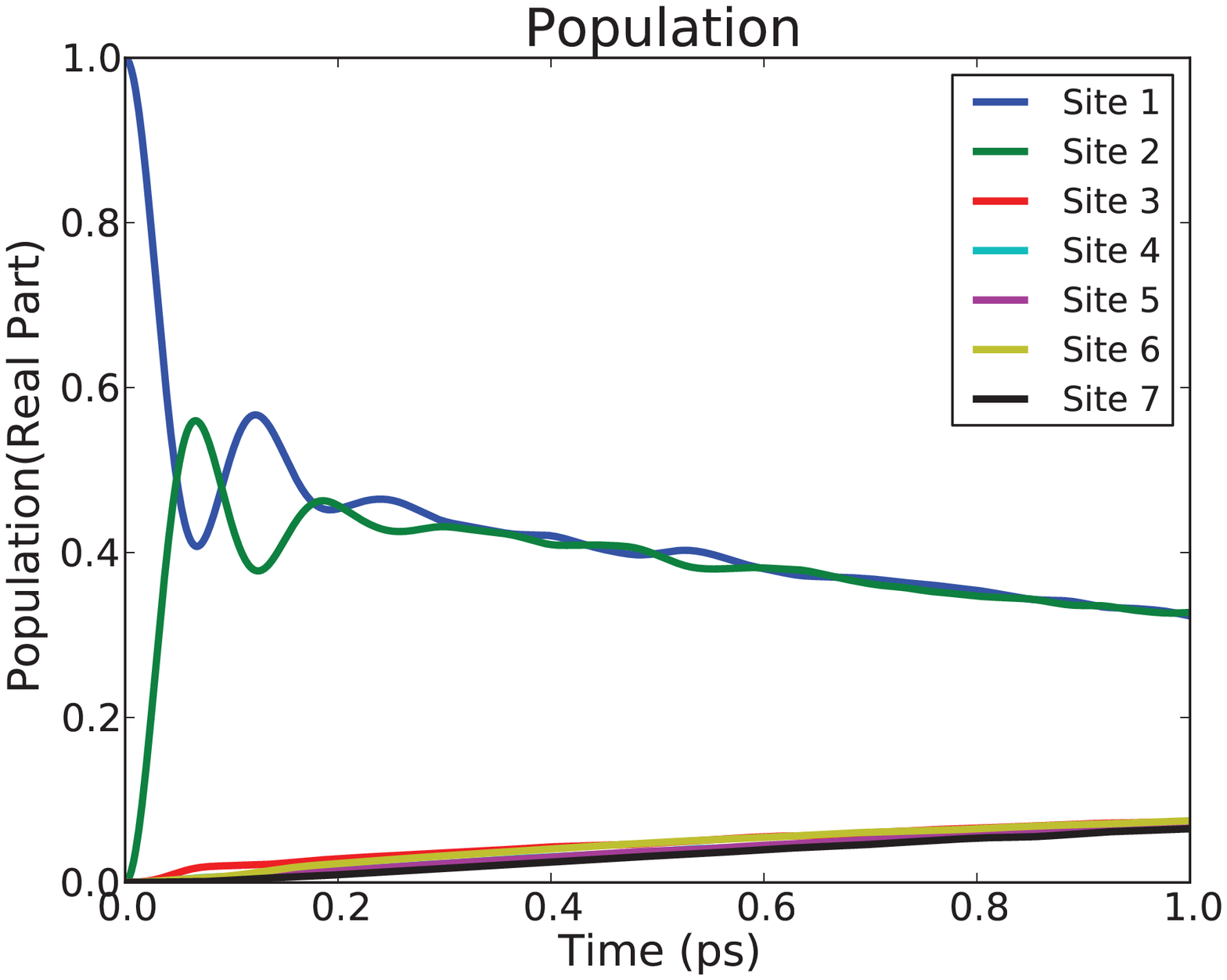}}\\
\end{tabular}
\caption{Population Dynamics of FMO, $\rho_{ii}(t)$, in the site representation for the kernels of four different spectral densities:  Lorentzian, Exponential Cutoff, and Gaussians with two different $\sigma$'s. (a) shows the results for Lorentzian, (b) the results for the kernel of exponential cut, (c) and (d) the results of kernels of Guassians with two different widths, $\sigma=\gamma$ and $\sigma=2\gamma$ ($\gamma=5\ ps^{-1}$)}\label{rpop}
\end{figure}

\begin{figure}
\begin{tabular}{cc}
\subfloat[]{\includegraphics[width=3in]{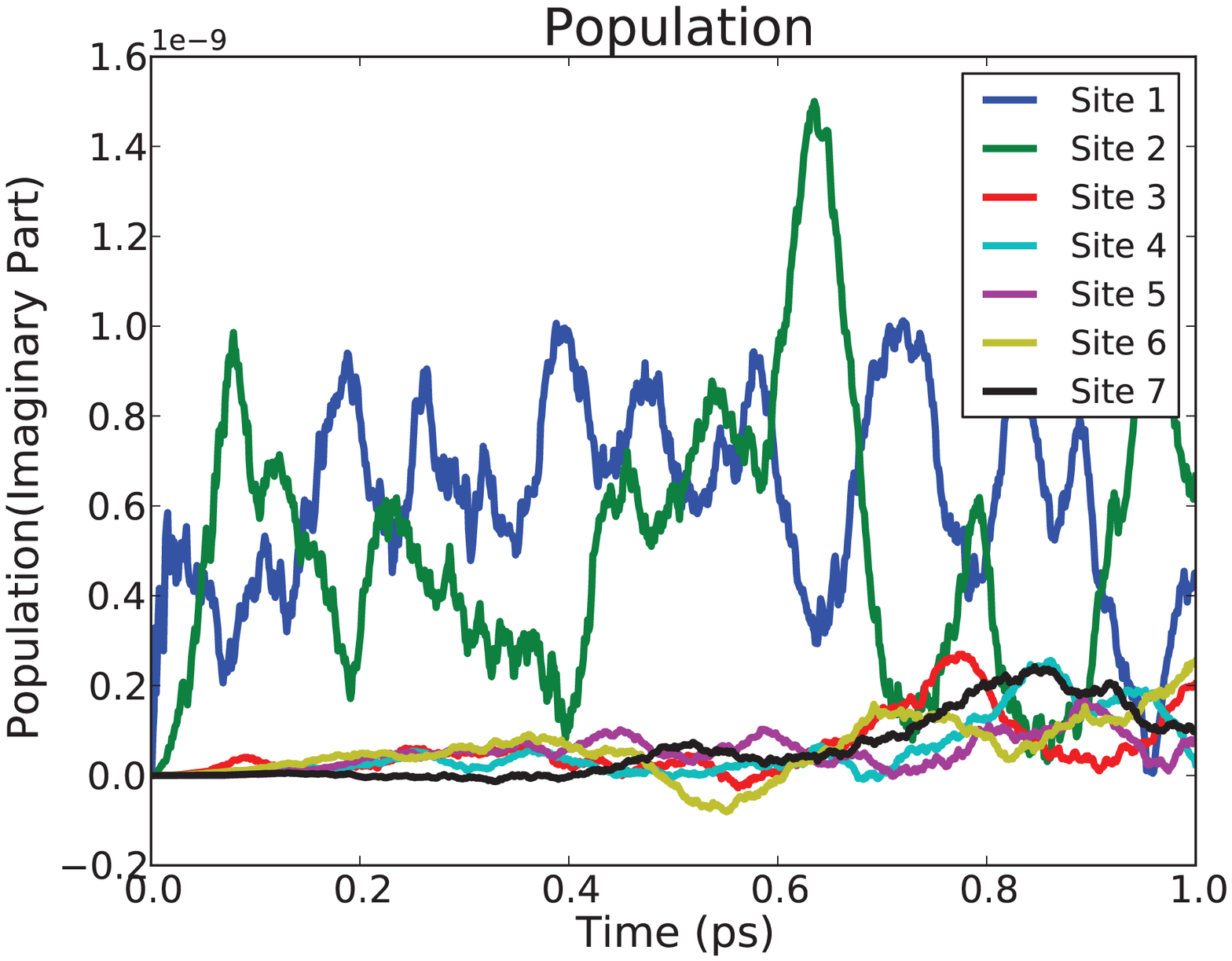}}
   & \subfloat[]{\includegraphics[width=3in]{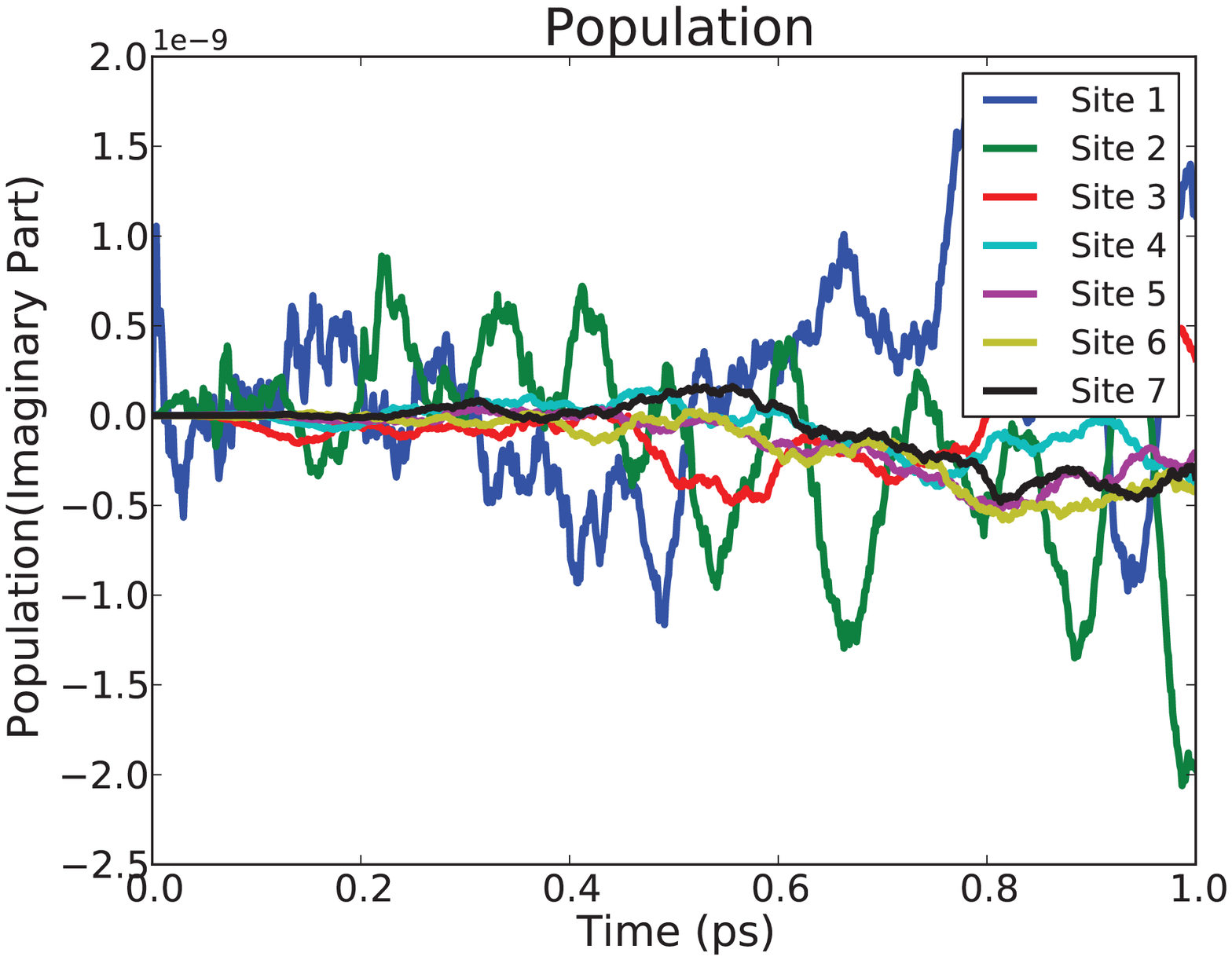}}\\
\subfloat[]{\includegraphics[width=3in]{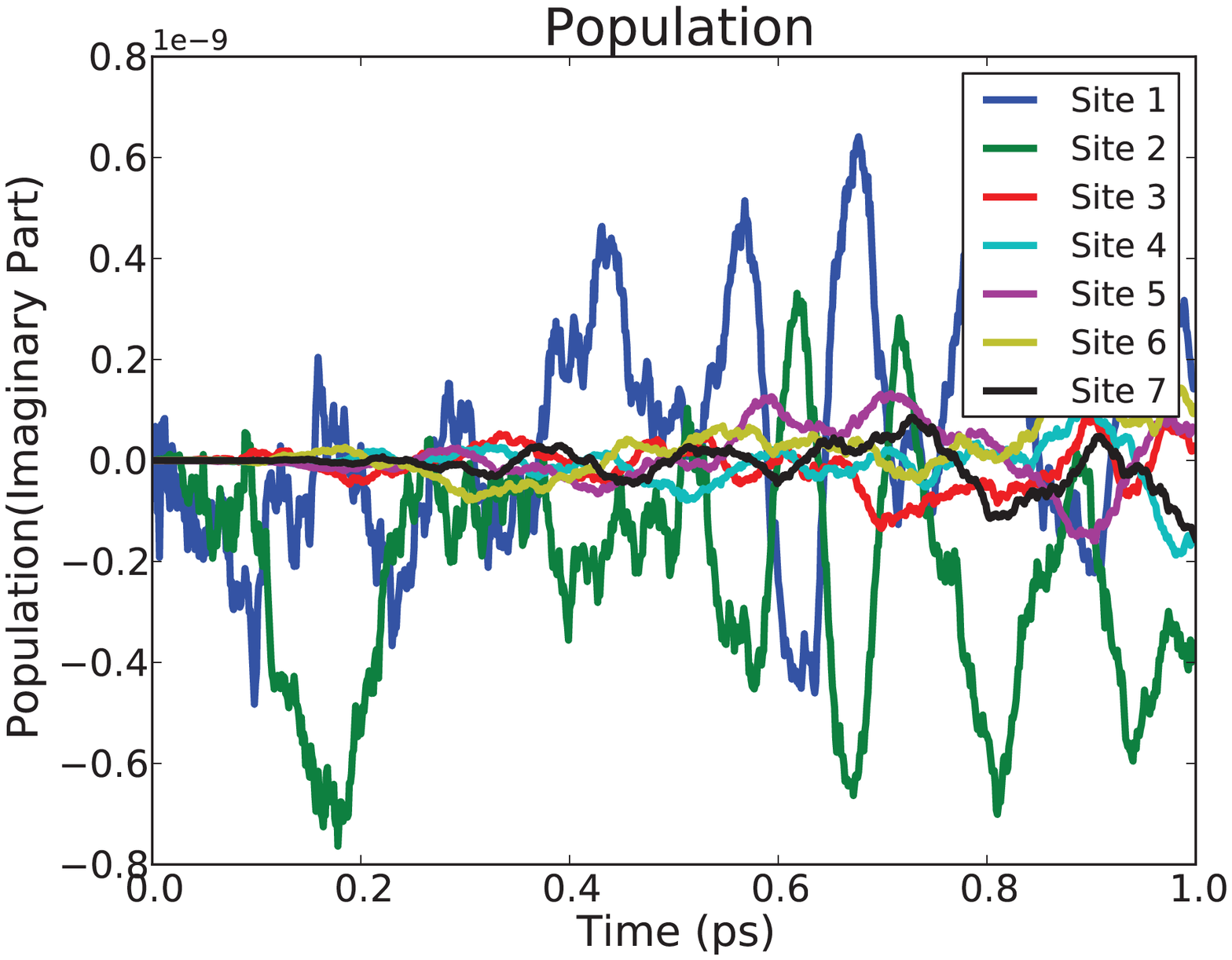}}
   & \subfloat[]{\includegraphics[width=3in]{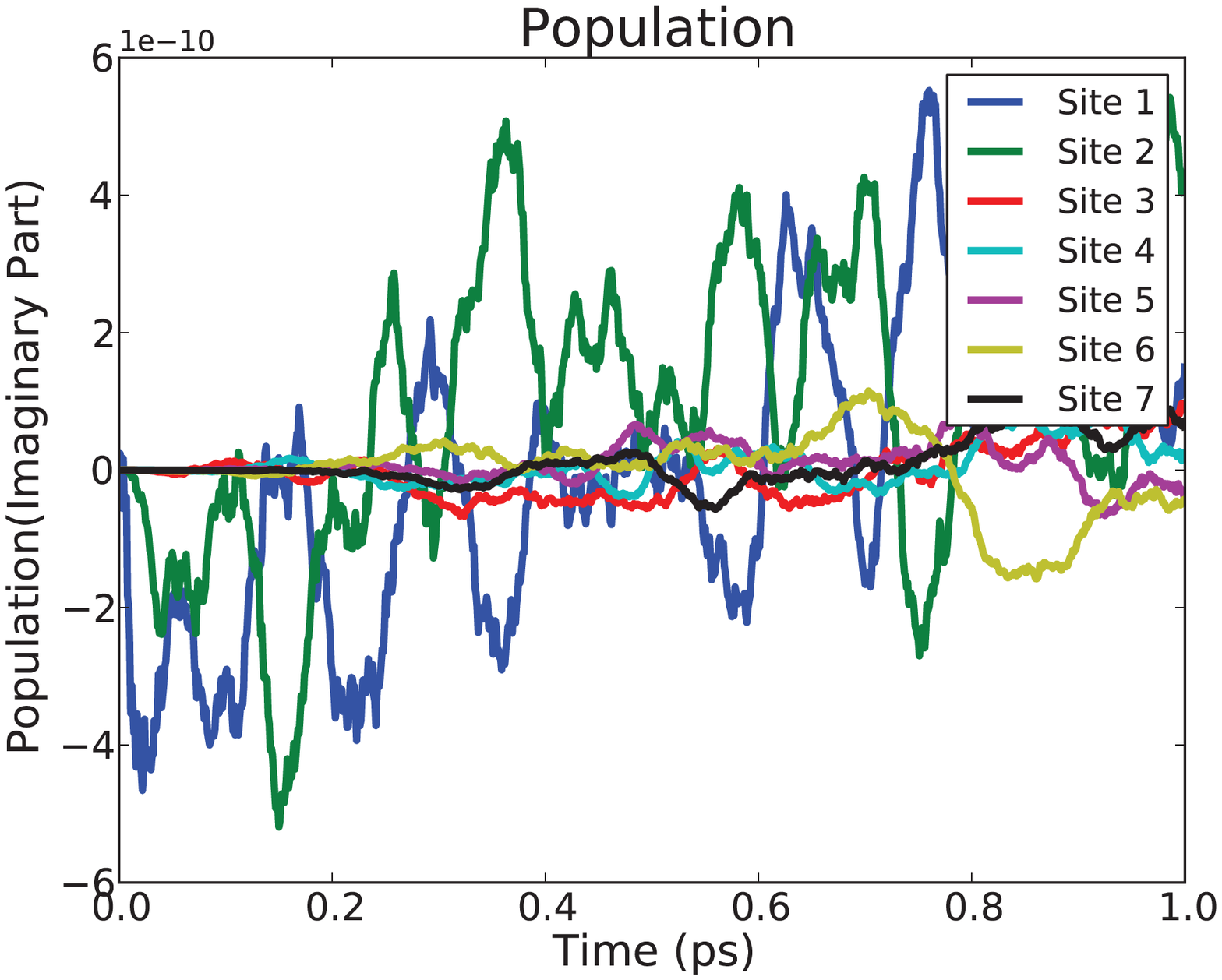}}\\
\end{tabular}
\caption{Corresponding imaginary part (in reference to Fig~\ref{rpop}) of the populations, $\rho_{ii}(t)$ (y-axis scale is at the magnitude of 1.0e-9 or 1.0e-10 for panel (d)) }\label{ipop}
\end{figure}

Fig~\ref{rpop} shows population dynamics $\rho_{ii}(t)$ for the four different spectral densities. Fig~\ref{ipop} clearly shows the imaginary parts of populations converge to zero. We can see that the Drude spectral density and second Gaussian spectral density give the fastest decay since both have more weight in the high frequency domain and smaller area under the real part of Kernel curve in Figure~\ref{kernel}.
Our conditional propagation scheme uses 5000 sampling trajectories for all the FMO results.

We also can see that while the Drude and first Gaussian spectral densities have the same peak, they have different areas under kernel curves and different decay rate.  Based on these observations, we can draw a simple conclusion that the low frequencies lead to the non-Markovian effect (memory effect) and high frequencies are more associated damping and decay\cite{reich}. But the areas under the kernel curves essentially decide decay/damping rates. However, the connection between spectral density and the area under the kernel curve is nonlinear. It suggests that tweaking with the shape of spectral density give us more interesting insight to the interplay between coherent and incoherent motions.

\begin{figure}
\centering
\begin{tabular}{cc}
\subfloat[]{\includegraphics[width=3.2in]{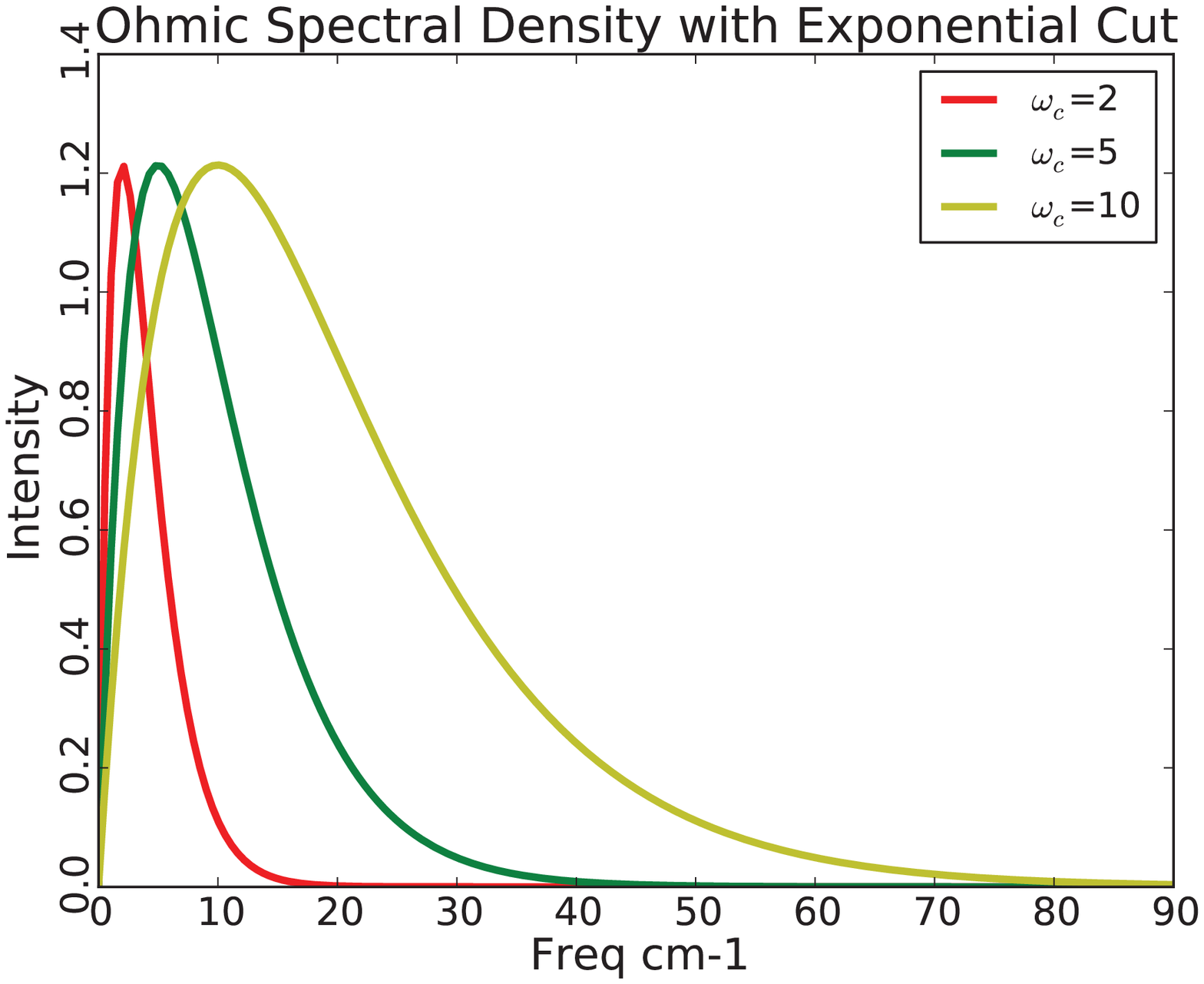}}
&\subfloat[]{\includegraphics[width=3.2in]{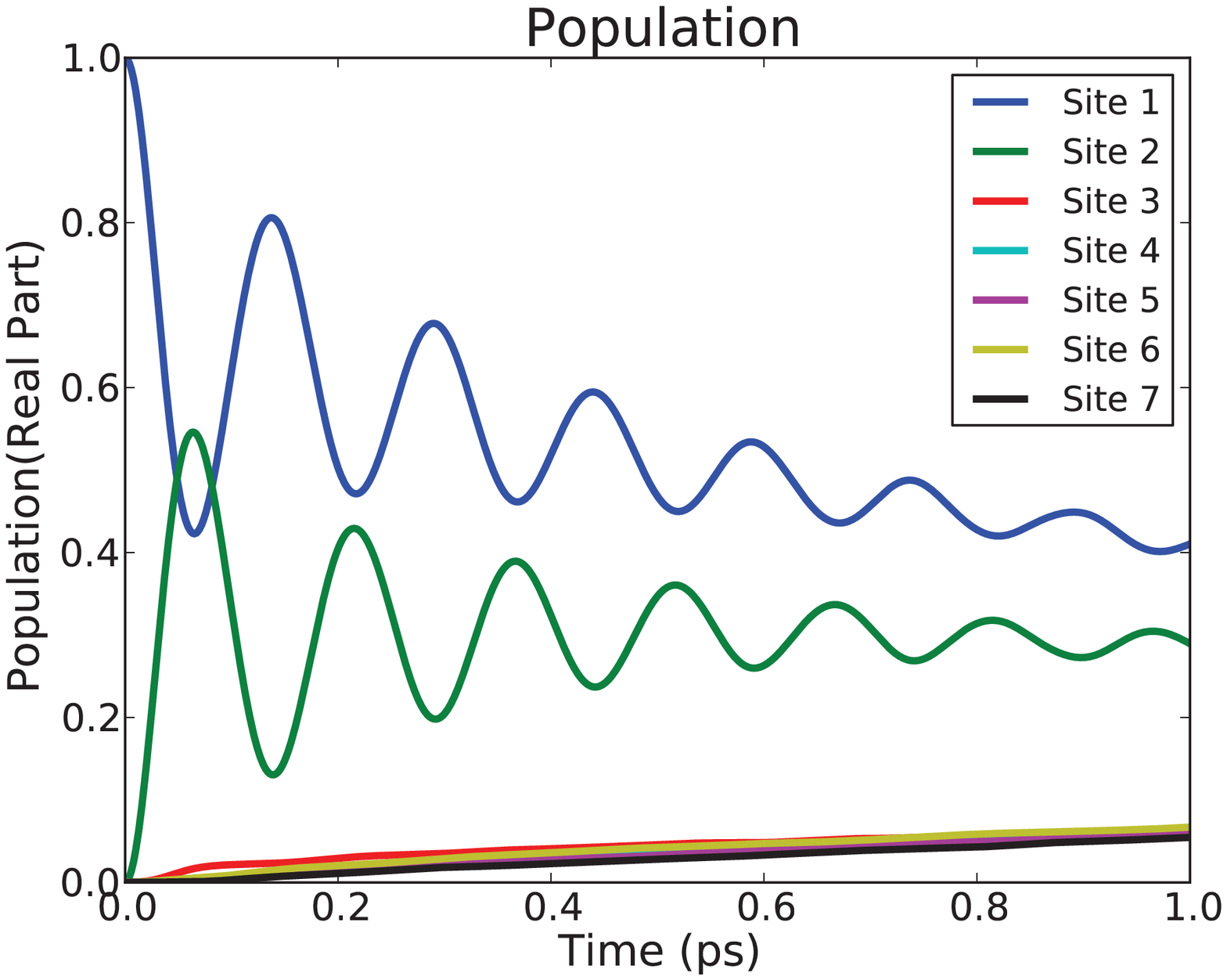}}\\
  \subfloat[]{\includegraphics[width=3.2in]{FMO-gec-pop-real.eps}}
 & \subfloat[]{\includegraphics[width=3.2in]{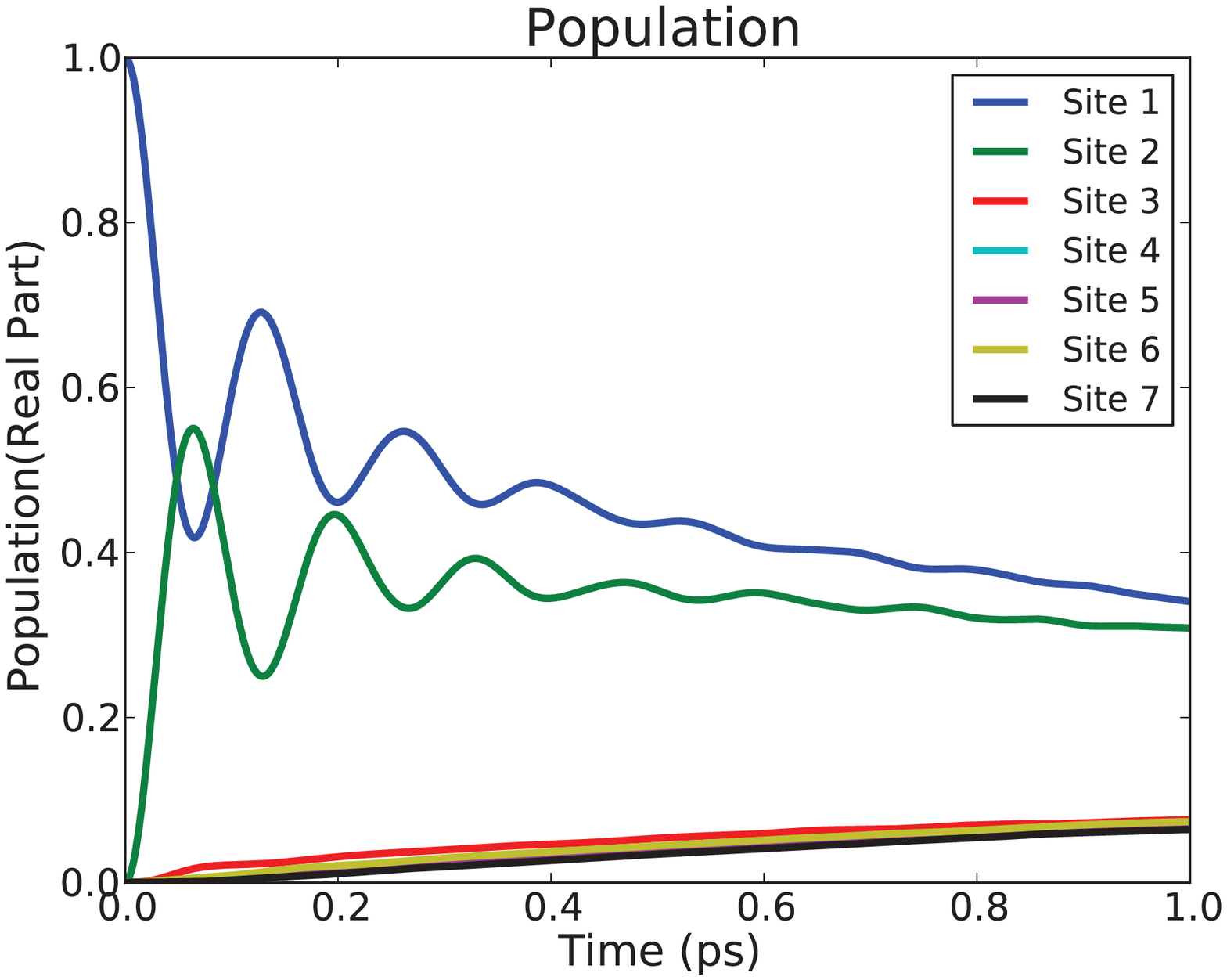}} \\
\end{tabular}
\caption{Panel (a) shows the spectral densities with differential exponential cutoffs, $\omega_c=2,5,10\ ps^{-1}$ and Panels (b), (c), and (d) show the comparison of the results, the population dynamics of reduced systems, $\rho_{ii}$, for the three different spectral densities.}\label{expcomp}
\end{figure}

Figure~\ref{expcomp} shows that the comparisons of the population dynamics for three different  Ohmic spectral densities with three different exponential cutoffs, $\omega_c=2,5,10\ ps^{-1}$. It is clearly shown that by fixing the reorganization energy, the more the spectral density is shifted to the higher frequency, the faster the relaxation/decay is.

In summary, based on the model we have here, the low frequencies is associated the memory (non-Markovian) effect and high frequencies are more associated with damping or decay. When we change the weight among low and high frequencies, we can change the curve of kernel and the relaxation process. All these arguments are based on the fixed initial conditions.

\section{Concluding Remark}\label{conclusion}

In the paper, we review coherent state path integral and influence functional. By exploiting the similarity between GCF and influence functional, we can construct complex-valued Gaussian processes to deconvolute the dynamics of reduced system temporarily. In the construction, we build the covariance matrix of PDF of discrete Gaussian propose according to convolution kernel of influence functional. Using the CD method, we can sample the discrete Gaussian process. On top of it, we propose conditional propagation scheme to simulate dynamics of the reduced system under the influence of different spectral densities.

Using the CD method and conditional propagation scheme, we examine EET in the FMO  complex under the influence of different spectral densities. We find that the geometry of spectral density can change the
reduced system dynamics. In this paper, we find that the the low frequencies is more associated with the non-Markovian effect (memory effect) and high frequencies are more associated damping and decay. Since the connection between spectral density and the area under the kernel curve is nonlinear, tweaking with the shape of spectral density may give us more interesting insight to the interplay between coherent and incoherent motions.

For the future work, we will extend the CD method to sample complex-valued Gaussian process according to the complex-valued non-Hermitian covariance matrix. We also like to apply the method to study of the correlated initial condition for the open quantum system.

\section{Acknowledgment}
This work was sponsored by the NSF (Grant No. CHE-1112825) and DARPA (Grant No. N99001-10-1-4063). The authors are partly supported by the Center for Excitonics, an Energy Frontier Research Center funded by the US Department of Energy, Office of Science, Office of Basic Energy Sciences under Award No. DE-SC0001088.

\appendix
\section{DGP and GCF} \label{dgp}
This representation of the stochastic DGP, $\xi(t)$ is characterized by its kernel, $\it{i.e.}$ two time correlation function,  $\gamma(t)=\langle \xi(t) \xi(0)\rangle$. In this section, we would like to show how we can construct the proper Gaussian process by treating the influence functional as a GCF.
Physicists are more familiar with the Fokker-Plank equation when we discuss the Brownian motion. From the practical perspective, Gaussian fluctuation's dynamic properties are determined by the correlation. Most of dynamical systems that we meet are stationary, which means the two-time correlation function is only dependent on the time difference, $t_i-t_{i-1}$.

DGP can be defined as a random vector $\hat{\xi}$ on the discrete time lattice, $[t_0,\cdots t_n]$, defined as,	
\begin{equation} \label{ranvec}
\hat{\xi}= [{\xi(t_0), \xi(t_1),\cdots, \xi(t_{n-1}), \xi(t_n)}]^T,
 \end{equation}
where $t_0=0$ and $t_n=t$. For the simplification of notations, we replace $\xi(t_i)$ with $\xi_{i}$.  The joint probability density function (PDF) of DGP $\hat{\xi}$ is a multivariate Gaussian
function. To define the multivariate Gaussian PDF, we need
two structure parameters, mean vector, $\hat{m}=[m_{1},m_{2},\cdots,m_{n}]^\dagger$ where $m_{i}=\langle \xi_{i}\rangle$, and covariance matrix, $\boldsymbol{\gamma}=[\gamma_{ij}]$ where $\gamma_{ij}=\langle(\xi_{i}-m_{i})(\xi_{j}-m_{j})\rangle$. Based on the mathematical properties of multivariate Gaussian function, PDF of DGP is defined as,
\begin{equation}\label{probden}
P(\hat{\xi})=N\exp[-(\hat{\xi}-\hat{m})^{\dagger}\boldsymbol{\gamma}^{-1}(\hat{\xi}-\hat{m})],
\end{equation}
where
\begin{equation}\label{covmat}
\boldsymbol{\gamma}=\left[\begin{array}{cccccc}
\langle(\xi_{0}-m_0)(\xi_{0}-m_0)\rangle & \langle(\xi_{0}-m_0)(\xi_{1}-m_1)\rangle & \cdots \\
\langle(\xi_{1}-m_1)(\xi_{0}-m_0)\rangle & \langle(\xi_{1}-m_1)(\xi_{1}-m_1)\rangle & \cdots \\
\vdots & \vdots & \ddots
\end{array}\right],
\end{equation} and
$N=\frac{1}{(2\pi)^{n/2}\vert \boldsymbol{\gamma} \vert^{1/2}}$
the normalization factor and $\vert \cdot \vert$ determinant. Without loss
of generality,
we assume that $\hat{m}=0$ in the context of mean field.

The Gaussian process $\xi(t)$ is the continuous limit to the
DGP, $\hat{\xi}$
when $dt\rightarrow0$. What is the meaning of the limitation? How do we express the limitation in terms of what quantity.
PDF of DGP $\hat{\xi}$  governs the discrete Gaussian environmental fluctuations. However, PDF's for continuous Gaussian processes don't exist. In other words, we can't take the continuous limit on PDF.
The Fourier transform of PDF's, $\it{i.e.}$ GCF, provides the equivalent information to PDF's. We can use GCF as the quantity to define the continuous limitation of DGP. In the following subsection, we will define the GCF and show the connection of GCF to influence functional.

Corresponding to Eq.~\ref{probden}, the GCF of DGO is defined as,
\begin{equation} \label{chdi}
\langle\exp(-i\;\sum_{i}\hat{s}^{T}\hat{\xi}\; dt)\rangle_{\hat{\xi}}=\exp(-\frac{1}{2}\sum_{ij}\hat{s}^{T} \gamma  \hat{s}\; dt^{2}),
\end{equation}
where we assume homogeneous time lattice, $t_i-t_{i-1}=dt$, and $\hat{s}=[s(t_{0}),s(t_{1}),s(t_{2}),\cdots,s(t_{n})]^\dagger$, the dummy vector (a discrete deterministic process). This is essentially the average of the Fourier Transform of PDF. By taking $dt\rightarrow0$ in Eq.~\ref{chdi},
the integral sign, $\int$, will replace the sum sign, $\sum$. Given the symmetry of the kernel, $\gamma(-t)=\gamma(t)$, we have the continuous-time Gaussian process GCF,
\begin{equation}\label{chco}
\begin{array}{ccc}
\langle\exp(-i\;\int_{0}^{t}\hat{s}^{\dagger}\hat{\xi}dt)\rangle & = & \exp[-\frac{1}{2}\int_{0}^{t}d\tau\int_{0}^{t} d\sigma\; s(\tau)
\gamma (\tau,\sigma) s(\sigma)]\\
 & = & \exp[-\int_{0}^{t} d\tau  \int_{0}^{\tau}d\sigma\; s(\tau) \gamma (\tau,\sigma) s(\sigma)].
\end{array}
\end{equation}
The continuous Gaussian process kernel, $\gamma(\sigma,\tau)$ is the the continuous limit of the discrete covariance matrix $\boldsymbol{\gamma}$  in Eq.~\ref{covmat}.

GCF for real (continuous-time) Gaussian process will give us the lead to construct the Quantum Gaussian random field to reproduce the influence functional. The simple comparison already shows that  influence functional is similar to GCF. Therefore, constructing the proper covariance matrix, $\boldsymbol{\gamma}$, for the complex-valued Quantum Gaussian random field, we can achieve our goal to map the convolution of influence functional to a random field. We will show how to construct the mapping in the next appendix section.

\section{Influence Functional and Complex-Value Gaussian Process}\label{ifgcf}
In order to draw the linkage between influence functional and general characteristic function, we need to discretize influence function.
Given that influence function in Eq.~\ref{eq:inf2} is a time ordered double integral and the symmetry of the Kernel $\gamma(-t)=\gamma^\dagger (t) $, we have
\begin{eqnarray}
F(\boldsymbol{Q},\boldsymbol{Q}';t_f,t_i) &=  & \exp  \{ - \int^{t_f}_{t_i} d\tau \int^{\tau}_{t_i} d\sigma \; (V (\boldsymbol{Q}(\tau)) - V(\boldsymbol{Q}'(\tau))  \\ \nonumber
&&  [\gamma(\tau-\sigma) V(\boldsymbol{Q}(\sigma))-\gamma^\dagger (\tau-\sigma) V(\boldsymbol{Q}'(\sigma))]   \} \\ \nonumber
& = & \exp \{  \int^{t_f}_{t_i} d\tau \int^{\tau}_{t_i} d\sigma \; -V (\boldsymbol{Q}(\tau)) \gamma(\tau-\sigma) V(\boldsymbol{Q}(\sigma) \} \\ \nonumber
& + & \exp \{  \int^{t_f}_{t_i} d\tau \int^{\tau}_{t_i} d\sigma \; -V (\boldsymbol{Q}'(\tau)) \gamma^\dagger(\tau-\sigma) V(\boldsymbol{Q}'(\sigma) \} \\ \nonumber
& + & \exp \{  \int^{t_f}_{t_i} d\tau \int^{\tau}_{t_i} d\sigma \; + V (\boldsymbol{Q}'(\tau)) \gamma(\tau-\sigma) V(\boldsymbol{Q}(\sigma) \} \\ \nonumber
& + & \exp \{  \int^{t_f}_{t_i} d\tau \int^{\tau}_{t_i} d\sigma \; + V (\boldsymbol{Q}(\tau)) \gamma^\dagger(\tau-\sigma) V(\boldsymbol{Q}'(\sigma) \}
\end{eqnarray}
We define $N$ homogenous discrete time grid $t_i$ where $i={0, 1,2,\cdots,N}$, $t_0=t_i$, $t_N=t_f$, and $dt=(t_f-t_i)/N$, $V_i = V(\boldsymbol{Q}(t_i)) dt $, $V'_i = - V(\boldsymbol{Q'}(t_i)) dt $ and covariance matrix element, $\gamma_{ij}=\gamma(t_i-t_j)$, $\gamma^\dagger_{ij}=\gamma^\dagger(t_i-t_j)$.
The discrete version of $F(\boldsymbol{Q},\boldsymbol{Q}';t_f,t_i)$ can be defined as,
\begin{equation}\label{discver}
\exp \left(- \sum_{i=0}^N\sum_{j=0}^i V_i \gamma_{ij} V_j  -	\sum_{i=0}^N\sum_{j=0}^i V'_i \gamma_{ij}^\dagger V'_j
-\sum_{i=0}^N\sum_{j=0}^i V'_i \gamma_{ij} V_j - \sum_{i=0}^N\sum_{j=0}^i V_i \gamma_{ij}^\dagger V'_j \right)
\end{equation}

Following the constructive approach in Sec.~\ref{ranop}, the covariance matrix for discrete complex-value Gaussian process, $\hat{\boldsymbol{\xi}}(t)$, is defined as,
\begin{equation}\label{comp_cov}
\boldsymbol{\gamma}=\left[\begin{array}{cccccc}
\langle\xi_{0}\xi_{0} \rangle & \langle \xi_{0} \xi_{1} \rangle & \cdots & \langle\xi_{0}\xi'_{0}\rangle & \langle\xi_{0}\xi'_{1} \rangle& \cdots\\
\langle\xi_{1}\xi_{0} \rangle & \langle \xi_{1} \xi_{1} \rangle & \cdots & \langle\xi_{1}\xi'_{0}\rangle & \langle\xi_{1}\xi'_{1} \rangle & \cdots\\
\vdots & \vdots & \ddots & \vdots & \vdots & \vdots\\
\langle\xi'_{0}\xi_{0} \rangle & \langle\xi'_{0}\xi_{1}\rangle & \cdots & \langle\xi'_{0}\xi'_{0} \rangle & \langle\xi'_{0}\xi'_{1} \rangle & \cdots\\
\langle\xi'_{1}\xi_{0} \rangle & \langle\xi'_{1}\xi_{1}\rangle & \cdots & \langle\xi'_{1}\xi'_{0}\rangle & \langle\xi'_{1}\xi'_{1}\rangle & \cdots\\
\vdots & \vdots & \vdots & \vdots & \vdots & \ddots
\end{array}\right],
\end{equation}
where
\begin{equation}
\hat{\boldsymbol{\xi}}=\left[\begin{array}{c}
\xi_{0}\\
\xi_{1}\\
\vdots\\
\xi_{n}\\
\xi'_{0}\\
\xi'_{1}\\
\vdots\\
\xi'_{n}
\end{array}\right],
\end{equation}
where $\xi_i = \xi (t_i)$ and $\xi'_i =\xi'(t_i)$
and the means of $\xi_i$ and $\xi'_i$ are zeros,
$\langle \xi_i \xi_j \rangle = \langle \xi_j \xi_i \rangle = \gamma_{ij}$,
$\langle \xi'_i \xi'_j \rangle = \langle \xi'_j \xi'_i \rangle = \gamma^{\dagger}_{ij}$,
$\langle \xi_i \xi'_j \rangle = \langle \xi'_j \xi_i \rangle = \gamma^{\dagger}_{ij}$ where $i \ge j$ and
$\langle \xi_i \xi'_j \rangle = \langle \xi'_j \xi_i \rangle = \gamma_{ij}$ where $i \le j$. Also $\gamma_{ii}=\gamma^{\dagger}_{ii}$. Therefore $\langle \xi_i \xi_i \rangle = \langle \xi'_i \xi'_i \rangle = \langle \xi'_i \xi_i \rangle =
\langle \xi_i \xi'_i \rangle = \gamma_{ii}=\gamma^\dagger_{ii}$. This matrix is non-Hermitian which is one intrinsic property of quantum open systems.

The corresponding complex-valued PDF of $\hat{\boldsymbol{\xi}}$ can be expressed as,
\begin{equation} \label{complex_den}
P=N\exp(-\frac{1}{2}\hat{\boldsymbol{\xi}}^{T}\;\boldsymbol{\gamma}^{-1}\;\hat{\boldsymbol{\xi}}),
\end{equation}
where
$N$ is the normalization factor. The corresponding GCF is defined as,
\begin{equation}\label{comp_cf}
\langle\exp(-i\;\sum_{i}\hat{\boldsymbol{V}}^{T}\hat{\boldsymbol{\xi}} )\rangle_{\hat{\boldsymbol{\xi}}}=\exp(-\frac{1}{2}\hat{\boldsymbol{V}}^{T} \boldsymbol{\gamma}  \hat{\boldsymbol{V}}),
\end{equation}
where the dummy vector, \[
\hat{\boldsymbol{V}}=\left[\begin{array}{c}
V_{0}\\
V_{1}\\
\vdots\\
V_{N}\\
V'_{0}\\
V'_{1}\\
\vdots\\
V'_{n}
\end{array}\right].
\]
After comparing Eq.~\ref{comp_cf} and Eq.~\ref{discver}, we can found our construction of complex-value Gaussian noises generates GCF equal to influence functional for the discrete version.
In order to prove the equality, we define the follow two vectors $\hat{\boldsymbol{V}}_1$ and $\hat{\boldsymbol{V}}_2$ and four blocks $\boldsymbol{\gamma}_{11}$, $\boldsymbol{\gamma}_{22}$, $\boldsymbol{\gamma}_{12}$, and $\boldsymbol{\gamma}_{21}$:
\[
\hat{\boldsymbol{V}}_1=\left[\begin{array}{c}
V_{0}\\
V_{1}\\
\vdots\\
V_{N}\\
\end{array}\right],
\]
\[
\hat{\boldsymbol{V}}_2=\left[\begin{array}{c}
V'_{0}\\
V'_{1}\\
\vdots\\
V'_{n}
\end{array}\right],
\]
\begin{equation}
\boldsymbol{\gamma}_{11}=\left[\begin{array}{cccccc}
\langle\xi_{0}\xi_{0} \rangle & \langle \xi_{0} \xi_{1} \rangle & \cdots \\
\langle\xi_{1}\xi_{0} \rangle & \langle \xi_{1} \xi_{1} \rangle & \cdots \\
\vdots & \vdots & \ddots
\end{array}\right],
\boldsymbol{\gamma}_{22}=\left[\begin{array}{cccccc}
\langle\xi'_{0}\xi'_{0} \rangle & \langle \xi'_{0} \xi'_{1} \rangle & \cdots \\
\langle\xi'_{1}\xi'_{0} \rangle & \langle \xi'_{1} \xi'_{1} \rangle & \cdots \\
\vdots & \vdots & \ddots
\end{array}\right],
\end{equation}
\begin{equation}
\boldsymbol{\gamma}_{12}=\left[\begin{array}{cccccc}
\langle\xi_{0}\xi'_{0} \rangle & \langle \xi_{0} \xi'_{1} \rangle & \cdots \\
\langle\xi_{1}\xi'_{0} \rangle & \langle \xi_{1} \xi'_{1} \rangle & \cdots \\
\vdots & \vdots & \ddots
\end{array}\right],
\boldsymbol{\gamma}_{21}=\left[\begin{array}{cccccc}
\langle\xi'_{0}\xi_{0} \rangle & \langle \xi'_{0} \xi_{1} \rangle & \cdots \\
\langle\xi'_{1}\xi_{0} \rangle & \langle \xi'_{1} \xi_{1} \rangle & \cdots \\
\vdots & \vdots & \ddots
\end{array}\right].
\end{equation}
Eq.~\ref{comp_cf} can be re-written as,
\begin{equation}
\exp(-\frac{1}{2}\hat{\boldsymbol{V}}_1^{T} \boldsymbol{\gamma}_{11}  \hat{\boldsymbol{V}}_1
-\frac{1}{2}\hat{\boldsymbol{V}}_1^{T} \boldsymbol{\gamma}_{22}  \hat{\boldsymbol{V}}_2
-\frac{1}{2}\hat{\boldsymbol{V}}_2^{T} \boldsymbol{\gamma}_{21}  \hat{\boldsymbol{V}}_1
-\frac{1}{2}\hat{\boldsymbol{V}}_1^{T} \boldsymbol{\gamma}_{12}  \hat{\boldsymbol{V}}_2
)
\end{equation}
Since $\boldsymbol{\gamma}_{11}$ and $\boldsymbol{\gamma}_{22}$ are symmetric matrices according to their definitions,
$\exp(-\frac{1}{2} \hat{\boldsymbol{V}}_1^{T} \boldsymbol{\gamma}_{11} \boldsymbol{V}_1 )=\exp(-\sum_{i=0}^N\sum_{j=0}^i V_i \gamma_{ij} V_j)$
and
$\exp(-\frac{1}{2} \hat{\boldsymbol{V}}_2^{T} \boldsymbol{\gamma}_{22} \boldsymbol{V}_2 )=\exp(-\sum_{i=0}^N\sum_{j=0}^i V'_i \gamma^{\dagger}_{ij} V'_j)$. For the remaining two terms, we have to consider the following eqaulity,
\begin{eqnarray}
&&\exp \big (-\frac{1}{2} \hat{\boldsymbol{V}}_1^{T} \boldsymbol{\gamma}_{12}
\hat{\boldsymbol{V}}_2-\frac{1}{2}\hat{\boldsymbol{V}}_2^{T}
\boldsymbol{\gamma}_{21} \hat{\boldsymbol{V}}_1 \big) =  \\ \nonumber
&&\exp \big (-\frac{1}{2} \sum_{i=0}^N\sum_{j=0}^i V_1^i \boldsymbol{\gamma}_{12}^{ij} V_2^j-\frac{1}{2}\sum_{i=0}^N\sum_{j=i+1}^N V_1^i \boldsymbol{\gamma}_{12}^{ij} V_2^j \\ \nonumber
&& -\frac{1}{2}\sum_{j=0}^N\sum_{i=0}^j V_2^i \boldsymbol{\gamma}_{21}^{ij} V_1^j-\frac{1}{2}\sum_{j=0}^N\sum_{i=j+1}^N V_2^i \boldsymbol{\gamma}_{21}^{ij} V_1^j \big )=\\ \nonumber
&&\exp \big(-\frac{1}{2} \sum_{i=0}^N\sum_{j=0}^i V_1^i \boldsymbol{\gamma}_{12}^{ij} V_2^j-\frac{1}{2}\sum_{i=0}^N\sum_{j=i+1}^N V_1^i \boldsymbol{\gamma}_{12}^{ij} V_2^j \\ \nonumber
&& -\frac{1}{2}\sum_{j=0}^N\sum_{i=0}^j V_1^j \boldsymbol{\gamma}_{21}^{ij} V_2^i-\frac{1}{2}\sum_{j=0}^N\sum_{i=j+1}^N V_1^j \boldsymbol{\gamma}_{21}^{ij} V_2^i \big)   =
\\ \nonumber
&&\exp \big(-\frac{1}{2} \sum_{i=0}^N\sum_{j=0}^i V_1^i \boldsymbol{\gamma}_{12}^{ij} V_2^j-\frac{1}{2}\sum_{i=0}^N\sum_{j=i+1}^N V_1^i \boldsymbol{\gamma}_{12}^{ij} V_2^j \\ \nonumber
&& -\frac{1}{2}\sum_{i=0}^N\sum_{j=0}^i V_1^i \boldsymbol{\gamma}_{21}^{ji} V_2^j-\frac{1}{2}\sum_{i=0}^N\sum_{j=i+1}^N V_1^i \boldsymbol{\gamma}_{21}^{ji} V_2^j\big ).
\end{eqnarray}
If $\boldsymbol{\gamma}_{12}^{ij}=\boldsymbol{\gamma}_{21}^{ji}=\gamma^\dagger_{ij}$ when $i>j$
and
$\boldsymbol{\gamma}_{12}^{ij}=\boldsymbol{\gamma}_{21}^{ji}=\gamma_{ij}$ when $i<j$ which are satisfied in the covariance matrix Eq.~\ref{comp_cov},
the equality becomes,
\begin{eqnarray}
&&\exp \big(-\frac{1}{2} \sum_{i=0}^N\sum_{j=0}^i V_1^i \boldsymbol{\gamma}_{12}^{ij} V_2^j-\frac{1}{2}\sum_{i=0}^N\sum_{j=i+1}^N V_1^i \boldsymbol{\gamma}_{12}^{ij} V_2^j \\ \nonumber
&& -\frac{1}{2}\sum_{i=0}^N\sum_{j=0}^i V_1^i \boldsymbol{\gamma}_{21}^{ji} V_2^j-\frac{1}{2}\sum_{i=0}^N\sum_{j=i+1}^N V_1^i \boldsymbol{\gamma}_{21}^{ji} V_2^j\big)=\\ \nonumber
&& \exp\big(-\sum_{i=0}^N\sum_{j=0}^i V_1^i \boldsymbol{\gamma}_{12}^{ij} V_2^j-\sum_{i=0}^N\sum_{j=i+1}^N V_2^j \boldsymbol{\gamma}_{21}^{ji} V_1^i\big).
\end{eqnarray}
It is easily to show that $\sum_{i=0}^N\sum_{j=i+1}^N V_2^j \boldsymbol{\gamma}_{21}^{ji} V_1^i= \sum_{i=0}^N\sum_{j=0}^i V_2^i \boldsymbol{\gamma}_{21}^{ij} V_1^j$. With this, we complete our proof,
\begin{eqnarray}
& &\langle\exp(-i\;\sum_{i}\hat{\boldsymbol{V}}^{T}\hat{\boldsymbol{\xi}} )\rangle_{\hat{\boldsymbol{\xi}}} = \\ \nonumber
& &\exp \left(- \sum_{i=0}^N\sum_{j=0}^i V_i \gamma_{ij} V_j  -	\sum_{i=0}^N\sum_{j=0}^i V'_i \gamma_{ij}^\dagger V'_j
-\sum_{i=0}^N\sum_{j=0}^i V'_i \gamma_{ij} V_j - \sum_{i=0}^N\sum_{j=0}^i V_i \gamma_{ij}^\dagger V'_j \right) ,
\end{eqnarray}
because $V_1^i$ corresponds $V_i$ and $V_2^i$ corresponds to $V'_i$.
By taking the continuous limit of Eq.~\ref{comp_cf}, the corresponding continuous-time GCF should match the influence functional in Eq.~\ref{eq:inf2}. In other words, we prove the equality in Eq.~\ref{random_op}.

\section{Complex Gaussian and Classical Gaussian Noises} \label{ccgn}
At the classical high temperature,  the complex covariance matrix will become
\begin{equation}\label{ht_cov}
\boldsymbol{\gamma}=\left[\begin{array}{cccccc}
\langle\xi_{0}\xi_{0} \rangle & \langle \xi_{0} \xi_{1} \rangle & \cdots & \langle\xi_{0}\xi'_{0}\rangle & \langle\xi_{0}\xi'_{1} \rangle& \cdots\\
\langle\xi_{1}\xi_{0} \rangle & \langle \xi_{1} \xi_{1} \rangle & \cdots & \langle\xi_{1}\xi'_{0}\rangle & \langle\xi_{1}\xi'_{1} \rangle & \cdots\\
\vdots & \vdots & \ddots & \vdots & \vdots & \vdots\\
\langle\xi'_{0}\xi_{0} \rangle & \langle\xi'_{0}\xi_{1}\rangle & \cdots & \langle\xi'_{0}\xi'_{0} \rangle & \langle\xi'_{0}\xi'_{1} \rangle & \cdots\\
\langle\xi'_{1}\xi_{0} \rangle & \langle\xi'_{1}\xi_{1}\rangle & \cdots & \langle\xi'_{1}\xi'_{0}\rangle & \langle\xi'_{1}\xi'_{1}\rangle & \cdots\\
\vdots & \vdots & \vdots & \vdots & \vdots & \ddots
\end{array}\right],
\end{equation}
which becomes a symmetric real-valued matrix, $\it{i.e.}$, the four matrices become equal to each other, $\boldsymbol{\gamma}_{11}=\boldsymbol{\gamma}_{22}=\boldsymbol{\gamma}_{21}=\boldsymbol{\gamma}_{12}$. In order to reproduce the high temperature real-valued covariance matrix, the random vector $\hat{\boldsymbol{\xi}}$ can be simplified to be,
\begin{equation}
\hat{\boldsymbol{\xi}}=\left[\begin{array}{c}
\xi_{0}\\
\xi_{1}\\
\vdots\\
\xi_{n}\\
\xi_{0}\\
\xi_{1}\\
\vdots\\
\xi_{n}
\end{array}\right],
\end{equation}
$\it{i.e.}$  $\xi'_i = \xi_i$.

\section{Gaussian Process for Dimers}\label{mulitlevel}
The influence functional can be defined as (for example in the reference\cite{makri1,makri2}),
\begin{eqnarray}
&&I(s_0^+, s_1^+, \dots,s,s_0^-,s_1^-,\dots,s') = \\ \nonumber
&&\text{Tr}_B[e^{-i (H_I (s, X) +I\times H_B(X) ) dt/2}e^{-i(H_I (s^+_{N-1}, X) + I\times H_B(X))dt}\dots \\ \nonumber
&& e^{-i (H_I (s^+_1, X)+I\times H_B(X)) dt}e^{-i (H_I (s^+_0, X) +I\times H_B(X)) dt} \rho_B (0) \\ \nonumber
&& e^{i (H_I (s^-_0, X)+I\times H_B(X)) dt} e^{i(H_I (s^-_{1}, X) +I\times H_B(X))dt}\dots \\ \nonumber
&&e^{i (H_I (s^-_{N-1}+I\times H_B(X)), X) dt}e^{i (H_I (s', X) +I\times H_B(X)) dt/2}],
\end{eqnarray}
where $s_i$ is the discrete system variables,
since
\begin{equation}
H_I=V\times X=\left( \begin{array}{cc}
H_{I1} & 0 \\
0 & H_{I2}
\end{array} \right)
\end{equation}
is diagonal, where $H_{I1}=V_1*X$ and $H_{I2}=V_2$ in the dimer Hamiltonian in Eq.~\ref{dimer_ham},
\begin{equation}
e^{-i(H_I (s_i, X) + I\times H_B(X)) dt} = 	
\left( \begin{array}{cc}
\exp (-i (V_1(s_i)*X +H_B(X)) dt)& 0 \\
0 & \exp( -i (V_2 (s_i)*X +H_B(X)) dt).
\end{array} \right)
\end{equation}
And the following influence functional,
\begin{eqnarray}\label{dimerinflu}
&&I(s_0^+, s_1^+, \dots,s,s_0^-,s_1^-,\dots,s') = \\ \nonumber
&&\text{Tr}_B \bigg [\left( \begin{array}{cc}
e^ {-i H_{e1}(s, X)  dt} e^{-iH_{e1} (s^+_{N-1}, X) dt}\dots e^{-iH_{e1} (s^+_0, X) dt} & 0 \\
0 & e^ {-i H_{e2}(s, X)  dt} e^{-iH_{e2} (s^+_{N-1}, X) dt}\dots e^{-iH_{e2} (s^+_0, X) dt} \\
\end{array} \right) \\ \nonumber
&&\times \rho_B(0) \\ \nonumber
&&\times \left( \begin{array}{cc}
e^ {i H_{e1}(s^-_0, X)  dt} \dots e^{iH_{e1} (s^-_{N-1}, X) dt} e^{iH_{e1} (s', X) dt} & 0 \\
0 & e^ {i H_{e2}(s^-_0, X)  dt} \dots e^{iH_{e2} (s^-_{N-1}, X) dt} e^{iH_{e2} (s', X) dt} \\
\end{array} \right) \bigg],
\end{eqnarray}
where $\rho_B(0)=\exp(-\beta H_B)/\text{Tr}(\exp(-\beta H_B))$, $H_{e1}=V_1(s_i)*X +H_B(X)$ and $H_{e2}=V_2(s_i)*X +H_B(X)$.
After integrating over the degree freedom of bath for the diagonal matrix elements in
Eq.~\ref{dimerinflu}, we get two separate influence functionals for each diagonal matrix element. As a result,
we should have two independent Gaussian random processes for each matrix elements.

\end{document}